\documentclass[pra,aps,twocolumn,showpacs]{revtex4-1}
\pdfoutput=1
\usepackage{dcolumn}
\usepackage{bm}

\usepackage[colorlinks,hyperindex, urlcolor=blue, linkcolor=blue,citecolor=black, linkbordercolor={.7 .8 .8}]{hyperref}
\usepackage{graphicx}
\usepackage{amsfonts}
\usepackage{amsmath}
\usepackage{amssymb}
\usepackage{amsbsy}
\usepackage{nicefrac}
\usepackage{color}
\usepackage{physics}

\newcommand{\rop}{\hat{a}^{\dagger}}
\newcommand{\lop}{\hat{a}}
\newcommand{\gnd}{|0\rangle}
\newcommand{\exc}{|1\rangle}
\newcommand{\fexc}{|2\rangle}

\providecommand{\abs}[1]{\lvert#1\rvert}

\usepackage{cancel}

\begin{document}

\title {Integrating superfluids with superconducting qubit systems}

\author{J.R. Lane$^1$}
\author{D. Tan$^{2,3}$}
\author{N.R. Beysengulov$^1$}
\author{K. Nasyedkin$^{1}$}
\author{E. Brook$^{1}$}
\author{L. Zhang$^{1}$}
\author{T. Stefanski$^{1}$}
\author{H. Byeon$^{1}$} 
\author{K.W. Murch$^{2,4}$} 
\author{J. Pollanen$^1$}
\email[]{pollanen@pa.msu.edu}
\affiliation{$^1$Department of Physics and Astronomy, Michigan State University, East Lansing, MI 48824-2320, USA}
\affiliation{$^2$Department of Physics, Washington University in St. Louis, St. Louis, MO 63130, USA}
\affiliation{$^3$Shenzhen Institute for Quantum Science and Engineering and Department of Physics, Southern University of Science and Technology, Shenzhen 518055, People's Republic of China}
\affiliation{$^4$Institute for Materials Science \& Engineering, Washington University in St.~Louis, 1 Brookings Dr., St.~Louis MO 63130, USA}

\date{\today}

\begin{abstract}

Superfluid helium's low-loss dielectric properties, excellent thermal conductivity, and  unique collective excitations make it an attractive candidate to incorporate into superconducting qubit systems. We controllably immerse a three-dimensional superconducting transmon qubit in superfluid $^{4}$He and measure the spectroscopic and coherence properties of the system. We find that the cavity, the qubit, and their coupling are all modified by the superfluid, which we analyze within the framework of circuit quantum electrodynamics (cQED). At at temperatures relevant to quantum computing experiments, the energy relaxation time of the qubit is not significantly changed by the presence of the superfluid, while the pure dephasing time modestly increases, which we attribute to improved thermalization of the microwave environment via the superfluid.

\end{abstract}

\maketitle

\section{Introduction}

Extensive progress in Josephson junction based superconducting qubits has made them a viable platform for building scalable quantum simulators and universal processors \cite{wen17, kri19, kra19}. In particular, the circuit quantum electrodynamic (cQED) architecture \cite{wal04,sch08}, wherein superconducting qubits are manipulated and read out with a superconducting microwave resonator, has been successful for implementing complex qubit control protocols and extending the coherence of quantum information stored in superconducting circuits. Coherence times $\gtrsim100~\mu$s are now routinely achieved in cQED experiments utilizing three dimensional (3D) resonators for both the information stored in the qubit \cite{pai11,rig12} and in the microwave resonator itself \cite{rea13,rea16,ofe16}. 

As quantum devices based on superconducting qubits become more sophisticated, challenges maintaining and extending the coherence required for useful applications have arisen. For example, microwave crosstalk and frequency crowding have been identified as limiting factors in current intermediate scale devices \cite{wen11,dev13}. The high-quality microwave environment afforded by 3D microwave cavities has been proposed as a way to mitigate these crosstalk issues \cite{bre16}, however the 3D cQED architecture is relatively inflexible in terms of frequency tuning. While SQUID tunable transmons may be employed to change the qubit frequency at the expense of additional flux noise, current methods of modifying the \textit{cavity} frequency, such as with a tuning screw, greatly deteriorate the electromagnetic quality \footnote{We note that \emph{two-dimensional} resonators can be frequency tuned, albeit with a reduced quality factor, via the incorporation of a SQUID loop in series with the resonator center conductor \cite{Pal08, San08}}. Additionally, 3D cQED experiments are known to suffer from poor thermalization of the microwave environment \cite{yeh17, yan18, wan19}, and from spurious excitations commonly associated with athermal superconducting quasiparticle poisoning \cite{mar09,cat112,pat17,ser18}. 

Given these considerations, controllably integrating superfluid helium into cQED experiments presents several potential advantages. For example, helium might serve as a thermalizing fluid to facilitate the cooling of the qubit and/or its environment. Helium immersion cells are already employed in the study of two-dimensional electron systems in the quantum Hall regime to more efficiently thermalize these systems at milli-Kelvin temperatures \cite{sam11,bra16}, however similar techniques have not been employed for superconducting qubit systems. It has also been shown that superfluid helium may be used to achieve several percent level changes in the resonant frequency of 3D microwave cavities \cite{sou17} similar to those used in cQED without significantly impacting the resonator's quality factor \cite{sou17,del14,del17}. While this amount of resonator frequency tuning is relatively small, as the number of superconducting qubits in quantum devices continues to scale, novel methods for fine frequency control could provide potential benefit in an ever more crowded frequency band \footnote{Additionally, recent experiments have demonstrated that superfluid helium can be used as the working fluid in a mechanically actuated 3D microwave cavity having a tunability $> 5$ GHz \cite{cla18}}.   

In addition to superconducting device considerations, there is also growing fundamental interest in studying the mechanical motion of superfluid helium at the quantum limit. Recent experiments and proposals have investigated the possibility of using superfluid helium as a platform for optomechanical experiments \cite{del14,del17,har16,chi17,sou17}, or as a substrate for an electron motional qubit \cite{pla99,dyk03,lyo06,schu10,yan162,nas18,koo19,bye19}, where details of the superfluid surface mechanics are important to understanding the decoherence of the proposed qubit. Additionally, rapid progress has recently been made using superconducting qubits to coherently control solid-state mechanical resonators at the several phonon level \cite{ocon10,moo18,sat18,chu18}. With these two classes of experiments in mind, it is natural to ask whether superconducting qubits could be employed to study or manipulate the mechanical or collective \cite{hal90,Dav08} excitations of quantum fluids. A prerequisite to any such experiment would be the characterization of the fundamental properties of a superfluid cavity/qubit coupled system, and a test to ensure that the presence of helium is not deleterious to the coherence properties of the qubit.

In this manuscript, we report on experiments studying the properties of a single-junction transmon \cite{koc07,pai11} superconducting circuit inside of a 3D microwave cavity resonator that can be controllably filled with superfluid $^{4}$He. We spectroscopically characterize the superfluid cavity/qubit system to determine the effect of the modified dielectric constant on the cavity, the qubit, and the coupling between the two. We also measure the decoherence properties of the qubit immersed in liquid helium. At the temperatures relevant to superconducting qubit experiments we find no significant degradation of the coherence properties of the qubit, and in fact observe a modest decrease in the pure dephasing rate in the presence of helium. We discuss these results in the context of possible relaxation and dephasing mechanisms and how they might be affected by the presence of the superfluid.

 
\section{Experimental setup}
\begin{figure}
\begin{center}
\includegraphics[width = 0.85\linewidth]{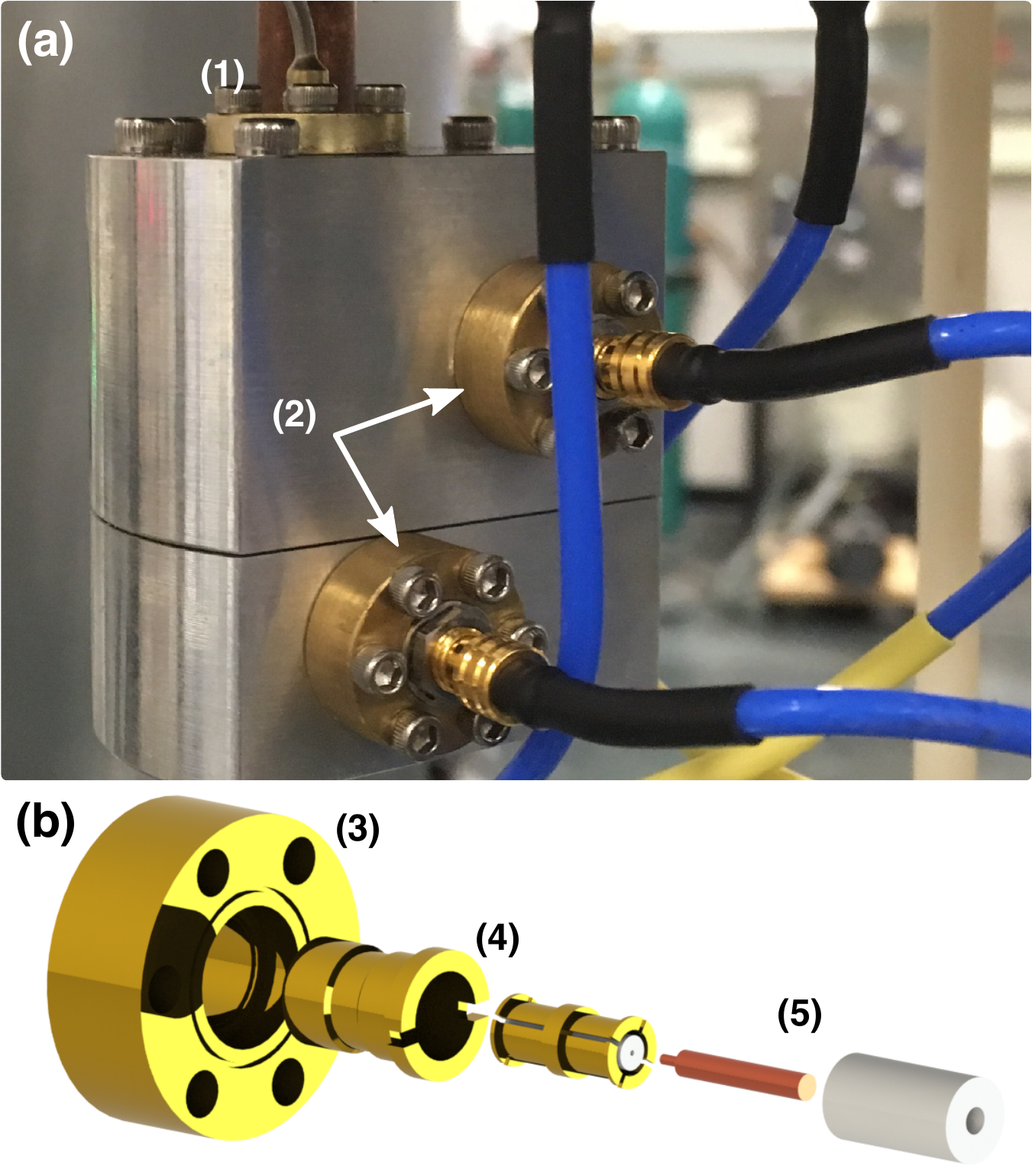}
\end{center}
\caption{(Color online) (a) Picture of the hermetically sealed 3D superconducting microwave cavity. Visible are (1) the helium fill capillary and flange and (2) the two microwave coupling ports. (b) Exploded rendering of the custom microwave coupling assembly. The hermetic GPO feedthrough (4) sits in a brass flange (3), and is sealed with an indium o-ring in between both the feedthrough and the flange, and the flange and the wall of the 3D cavity. A $50~\Omega$ impedance matched copper pin (5) is soldered into a GPO ``bullet'' connector and extends to the inner wall of the cavity and provides coupling to the TE101 fundamental mode of the cavity.}
\label{fig1}
\end{figure}

The experiment consists of a single-junction transmon circuit housed in a rectangular 3D aluminum microwave cavity (see Fig. $\ref{fig1}$(a)). The two halves of the cavity are hermetically sealed with a conventional indium wire o-ring typically used for making superfluid leak tight joints. The external microwave coupling to the cavity is provided via two hermetically sealed $50~\Omega$ assemblies (see Fig. $\ref{fig1}$(b)). These assemblies consist of commercial hermetic GPO feedthroughs \footnote{Gilbert Engineering part \# 0119-783-1} in which the room temperature rubber o-ring has been replaced with a cryogenic indium seal \cite{Fra13,nas18}. These GPO feedthroughs are seated in custom brass flanges that are themselves sealed to the body of the cavity with indium o-rings. Coupling pins are soldered to the inner portion of each assembly to provide microwave signals to the cavity. The microwave ports are connected to standard filter/amplifier chains used in cQED experiments (see Appendix A).

The cavity is thermally anchored to the mixing chamber plate of a cryogen-free dilution refrigerator, and helium is introduced via a hole on the top of the cavity that connects to a stainless steel fill capillary through a custom brass flange, which is itself hermetically sealed to the cell with an indium o-ring. This fill capillary extends the length of the cryostat and the helium is thermalized via copper sinter heat exchangers positioned at the still plate, cold plate, and mixing chamber of the dilution refrigerator. In this configuration, we find that the lowest temperature of the dilution refrigerator is not substantially changed from its nominal value of $10$~mK upon filling the 3D cavity with superfluid $^{4}$He.  

\section{Cavity and qubit spectroscopy}

\begin{figure}
\begin{center}
\includegraphics[width=1 \columnwidth]{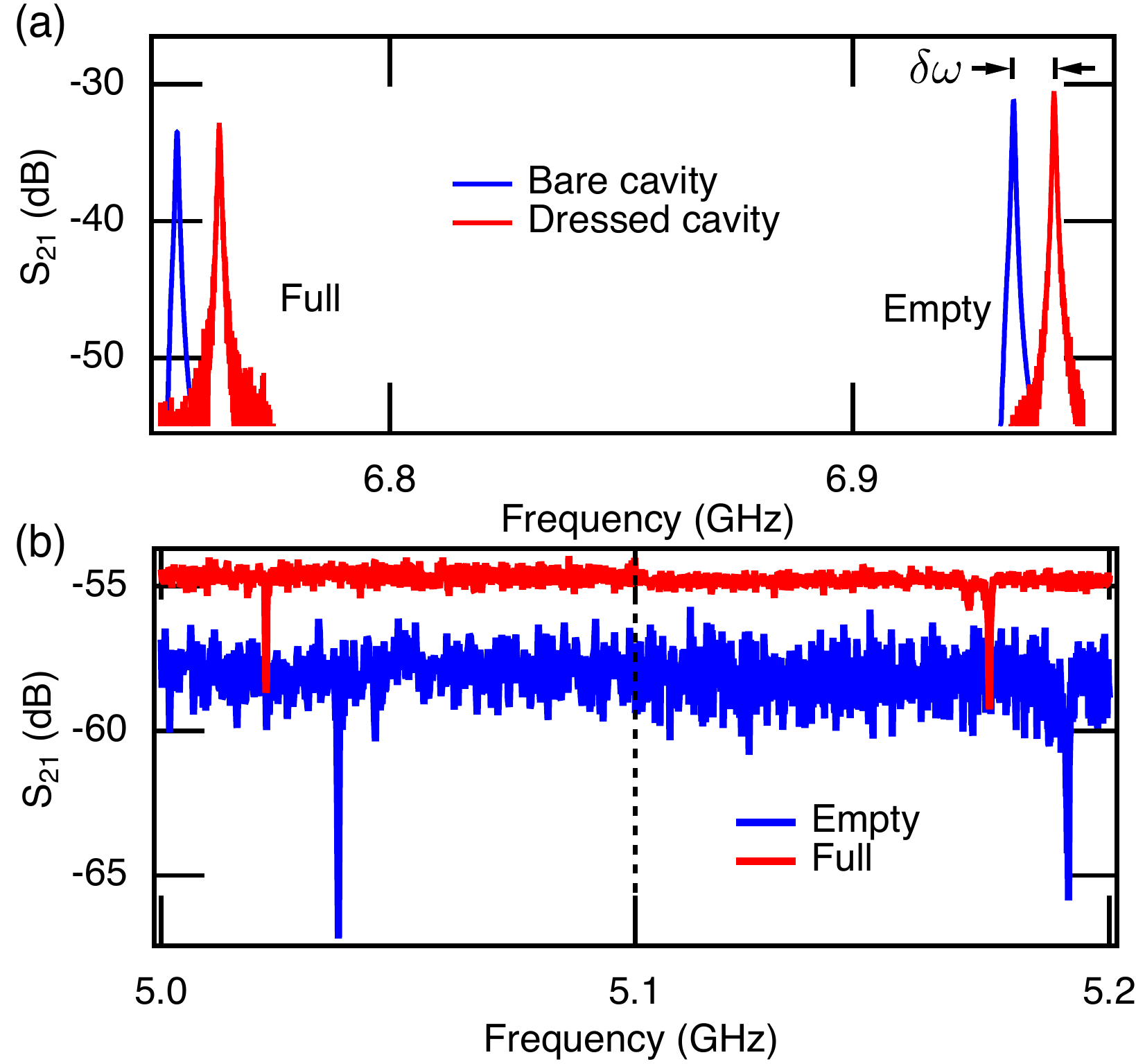}
\caption{(Color online) (a) Measured cavity transmission $S_{21}$ as a function of frequency when the cavity is empty (right) and full of superfluid helium (left). Depending on the level of the applied microwave power we can measure both the cavity resonance dressed by the qubit in its ground state (red, $P\approx-120$~dBm) or the bare cavity resonance (blue, $P\approx-80$~dBm.) (b) Two-tone spectroscopy of the qubit immersed in liquid helium (red) and in vacuum (blue), offset vertically for clarity. Right of the dotted line, a low power tone is applied to excite the qubit ($P_q\approx-120$~dBm) from its ground state $\gnd$ to its first excited state $\exc$, while to the left of the line a high power ($P_q\approx-90$~dBm) tone is applied to induce a two photon transition from $\gnd$ to $\fexc$. The dips in the transmission correspond to qubit excitation frequencies $\omega_{01}$ (right) and $(\omega_{01}+\omega_{12})/2$ (left).}
\label{fig2}
\end{center}
\end{figure}

To characterize the effect of adding superfluid $^{4}$He we first perform continuous wave spectroscopy of the cavity/qubit coupled system, both when the cavity is empty and under vacuum, and when it is filled with superfluid helium. Using a vector network analyzer we characterize the cavity response by measuring the microwave transmission ($S_{21}$) through the measurement circuit as a function of frequency. At high power ($\sim -80$~dBm power injected into the cavity), the measured response is Lorentzian and peaked at the classical cavity fundamental frequency \cite{boi10,bis10,ree10} $f_c=\omega_c/2\pi$, shown as the blue traces in Fig. $\ref{fig2}$(a). The change in the speed of light caused by the presence of a dielectric of relatively permittivity $\epsilon$ should shift the bare cavity frequency from $f_c \rightarrow f_c/\sqrt{\epsilon}$. Indeed, we find that, when helium is added to the cavity, the fundamental frequency $f_c$ shifts from 6.93480~GHz to 6.75395~GHz (see Table~\ref{table1}), corresponding to an effective cavity dielectric constant of $\epsilon = 1.054$, which agrees well with that of superfluid helium $\epsilon_{He} = 1.057$ \cite{sou17,bro77}. We also note that the quality factor of the microwave resonator is not significantly affected by the presence of helium, consistent with the findings in Ref. \cite{del14, del17, sou17}.

The hybrid cavity/qubit system is described by the generalized Jaynes-Cummings Hamiltonian (JCH), which takes into account the higher excited states of the transmon circuit $|i\rangle$:
\begin{equation}
\hat{H}_{JC} = \hbar\omega_c\rop\lop + \sum_i \omega_i |i\rangle \langle i| + \hbar \sum_i \big(g_{i,i+1} |i\rangle\langle i+1 | \rop + \text{h.c.} \big).
\end{equation}
In Eq.~1 $\rop$ and $\lop$ correspond to the microwave photon creation and annihilation operators respectively. In the transmon regime \cite{koc07}, the uncoupled qubit frequencies $\omega_i$ are determined by the Josephson energy $E_J$, the charging energy $E_C$, and the cavity/qubit couplings constants $g_{i,i+1} \approx g_{01}\sqrt{i+1}$. In this limit, Eq.~1 is therefore determined by $E_J$, $E_C$, the ground-to-first excited state vacuum Rabi splitting $g_{01}$, and the cavity frequency $\omega_c$.

In addition to shifting the cavity resonant frequency, the presence of dielectric superfluid will also modify all of the spectroscopic parameters of the coupled qubit/cavity system, which we can characterize with the framework of cQED. At low input microwave power ($\sim -120$~dBm, Fig. $\ref{fig2}$(a) red traces) the cavity resonant frequency is shifted by the presence of the transmon circuit in its ground state. In the dispersive limit of cQED \cite{koc07}, $|\Delta|=|\omega_{c}-\omega_{01}|\gg g_{01}$, where $\omega_{01}$ is the qubit ground-to-excited state frequency, this hybridization causes the cavity resonant frequency to shift by an amount $\delta \omega \approx g^2_{01}/\Delta$. We measure $\delta \omega$ for both the empty cavity and the cavity full of superfluid helium, with the results reported in Table~\ref{table1}.

We utilize two-tone spectroscopy \cite{schu05} to directly measure the excitation spectrum of the qubit and how it is modified by the superfluid. We use a low power tone (Fig. \ref{fig2} (b), right of dashed line) to excite the qubit from ground $\gnd$ to first excited state $\exc$, and a higher power tone to excite a two photon transition from $\gnd$ to the second excited state $\fexc$ (Fig. \ref{fig2} (b), left of dashed line). From these measurements, we extract the $\gnd \rightarrow \exc$ transition frequency $\omega_{01}$ and the $\exc \rightarrow \fexc$ transition frequency $\omega_{12}$ for both the empty and full cavity configurations, and report these values in Table~\ref{table1}.
\begin{table}
\begin{center}
\begin{tabular}{|c || c | c | c|} 
 \hline
 Value & Empty (GHz) & Full (GHz) & change (\%) \\ [0.5ex] 
 \hline
 $\omega_c/2\pi$ & 6.9348 & 6.7540  & -2.62 \\ 

 $\delta \omega/2\pi$ & 0.00875 & 0.00913 & 4.32 \\

 $\omega_{01}/2\pi$ & 5.1914 & 5.1747 & -0.32 \\
 
 $\omega_{12}/2\pi$ & 4.8834 & 4.8695 & -0.28 \\
 
 $E_J/h$ & 13.887 & 13.895 & 0.06 \\
 
 $E_C/h$ & 0.2710 & 0.2690 & -0.82 \\
 
 $g_{01}/2\pi$ & 0.1235 & 0.1201 & -2.8 \\[1ex] 
 \hline
\end{tabular}
\caption{Spectroscopic parameters of the cavity/qubit system both in the presence and absence of superfluid helium. $\omega_c, \delta \omega, \omega_{01}$, and $\omega_{12}$ are measured values, while $E_J, E_C$ and $g_{01}$ are extracted by solving the generalized Jaynes-Cummings Hamiltonian constrained by measured spectroscopic parameters.}
\label{table1}
\end{center}
\end{table}

To extract the values of $E_J$, $E_C$, and $g_{01}$, and how they are modified by the dielectric superfluid, we diagonalize the generalized JCH, and fit the eigenvalues $\omega_{01}$, $\omega_{12}$, and $\delta \omega$ to the values obtained from our spectroscopy measurements for the case when the 3D cavity is empty as well as when it is filled with helium. The results are is summarized in Table \ref{table1}.

The small change of $E_J$ in the presence of helium is consistent with variations in $E_J$ that we observe between cool downs without helium present in the cavity. It has been reported that these variations result from changes in the microscopic charge configuration in the Josephson junction oxide barrier \cite{rog85,van04}. Therefore our results are consistent with $E_J$ being unmodified by the presence of liquid helium. In contrast we find that the capacitive charging energy of the qubit decreases by $0.82\%$. This reduction in $E_C$ agrees with the value of $0.78\%$ obtained from finite element simulations of the system (see Appendix B).

A shift in the vacuum Rabi coupling $g_{01}$ is also induced by the superfluid helium. Qualitatively, this shift results from a change in the zero point energy of the cavity \emph{and} a spatial redistribution of electric field lines within the cavity/qubit system upon changing the dielectric constant from $\epsilon = 1 \rightarrow \epsilon_{He} = 1.057$. Quantitatively, we write the vacuum Rabi coupling in terms of the fluctuating zero point voltage of the microwave field in the 3D cavity \cite{koc07} $V_{ZPF}$,
\begin{equation}
g_{01} = 2eV_{ZPF} \beta \langle1\vert \hat{n} \vert 0 \rangle,
\end{equation}
where $\hat{n}$ is the Cooper pair number operator, and $\beta$ is a parameter describing the efficiency of converting voltage fluctuations in the cavity to voltage fluctuations across the junction of the qubit \cite{koc07}. We develop a simple model (see Appendix B) describing how $\beta$ depends on the dielectric constant of the cavity. This model, when taken into account with the dielectric induced modifications in $V_{ZPF} \propto \epsilon^{-3/4}$ and $ \langle1\vert \hat{n} \vert 0 \rangle \propto E_C^{-1/4}$, yields a predicted shift in the vacuum Rabi splitting $\Delta g_{01} =-3.2\%$ in comparison with the measured shift of $\Delta g_{01} = -2.8\%$ produced by the superfluid (see Table~\ref{table1}), in good agreement given the relatively simple circuit model employed in our analysis.

\section{Qubit Relaxation and Decoherence}

\begin{figure}
\begin{center}
\includegraphics[width=1 \columnwidth]{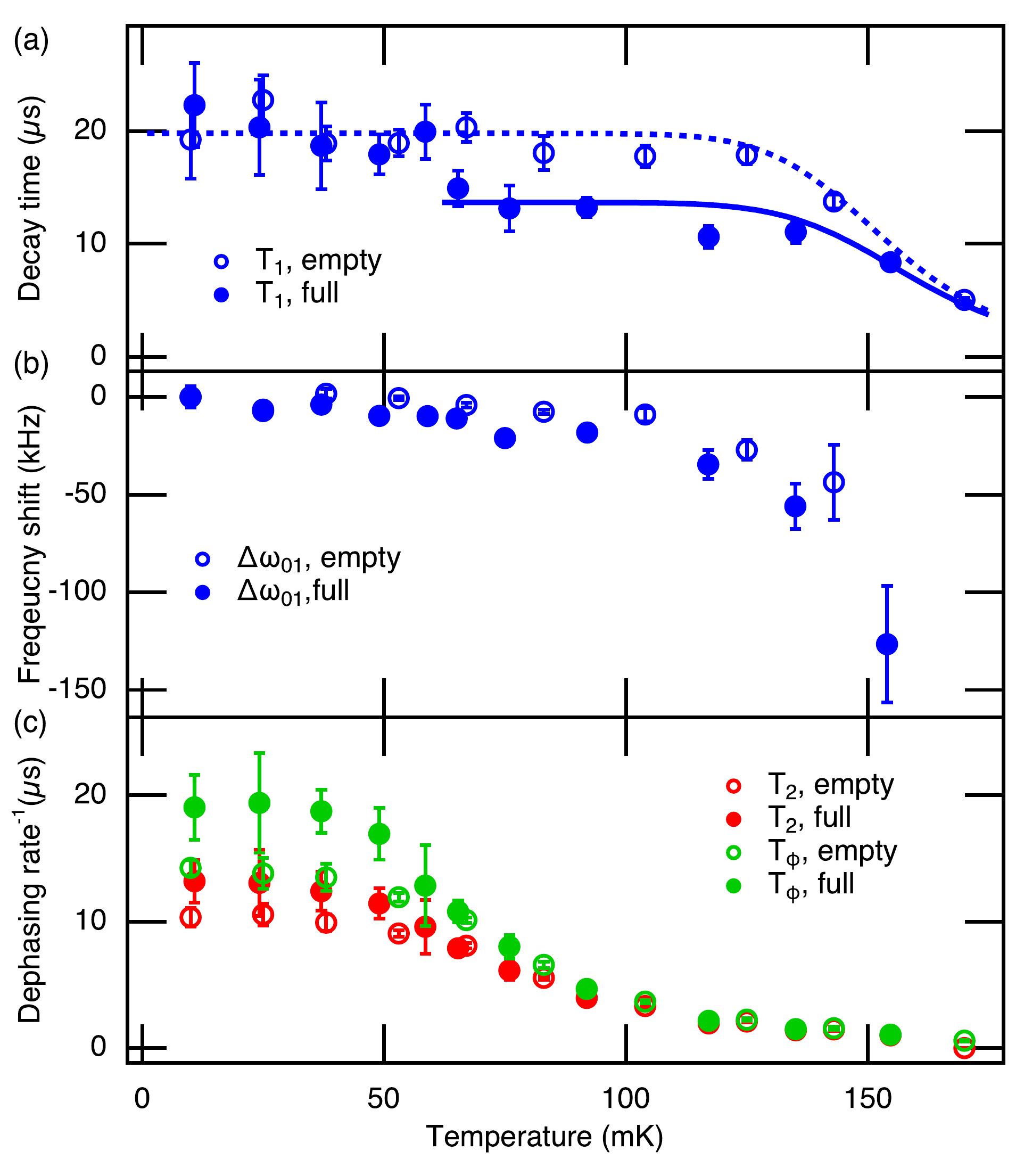}
\caption{(Color online) (a) Qubit energy relaxation time $T_1$, as a function of temperature, for both the empty cavity and the cavity filled with superfluid helium. We fit the data to theory (see Ref. \cite{cat112}) for quasiparticle limited $T_1$, using only the superconducting gap of aluminum and the nonequilibrium quasiparticle density $x_{\mathrm{qp}}$ (solid and dashed curves) to extract the change in quasiparticle density above $\sim 60$ mK when the cavity is filled with superfluid (see main text for discussion.) 
(b) Qubit frequency shift relative to it's base temperature value, (c) dephasing ($T_2$) and pure dephasing ($T_\phi$) times of the qubit as a function of temperature, for both the empty and full cavity configurations.}
\label{fig3}
\end{center}
\end{figure}

Successful integration of qubits into quantum fluid experiments or vice vera requires that the coherence properties of the qubit do not degrade when immersed in superfluid helium. To characterize these effects, we use standard pulsed techniques to measure the energy ($T_1$) and phase ($T_2$) relaxation of the qubit as a function of temperature. The temperature dependences of $T_{1}$ (blue) and $T_{2}$ (red) are plotted in Fig.~\ref{fig3} for the case when the cavity is empty (open symbols) and when it is filled with helium (closed symbols). We repeatedly measure $T_{1}$ and $T_{2}$ at each temperature for 5 hours to average over long timescale fluctuations (for a discussion of these long time scale fluctuations see Appendix C.) The data points in Fig.~\ref{fig3} represent the average value of a set of repeated measurements, while the error bars are the standard deviation of each set. From these values of $T_{1}$ and $T_{2}$ we also calculate the pure dephasing time $T_\phi~=~(1/T_2~-~1/2T_1)^{-1}$, which are shown as the green symbols in Fig.~\ref{fig3}(c).

\subsection{Qubit energy relaxation}

At the lowest temperatures we find that $T_1$ saturates at roughly the same value ($\sim20~\mu$s) both when the cavity is empty and when it is full of superfluid helium. Whatever mechanism is limiting $T_1$ at the lowest temperature we can conclude that it is not significantly suppressed by the presence of superfluid helium \footnote{We note that at the large qubit/cavity detuning used in this experiment, the increase in the Purcell emission rate $\Gamma_p$ caused by the helium induced shift of the cavity frequency is negligible compared to the long timescale fluctuations in the emission rate. Specifically $\Gamma_p = (g_{01}/\Delta)^2 \kappa\sim(265~\mu$s$)^{-1}$ for the empty cavity and $\sim(240~\mu$s$)^{-1}$ for the full cavity, where $\kappa$ is the cavity linewidth \cite{koc07}.}. On the other hand, this result also demonstrates that the superfluid does not introduce any significant additional mechanisms for qubit energy relaxation at the temperatures relevant to cQED. 

The temperature dependence of $T_1$ when the cavity is not filled with helium (see Fig.~\ref{fig3}(a)) may be understood as arising from quasiparticles tunneling across the qubit junction \cite{mar09,cat112,jin15,ser18}, limited by an athermal quasiparticle bath below $\sim140$~mK. However, when the cavity is filled with helium we observe a qualitatively different temperature dependence: as we increase the temperature above $\sim 60~$mK, we find a modest reduction in $T_1$ when the cavity is filled with superfluid. We posit that this reduction in $T_1$ could be associated with a higher nonequilibrium quasiparticle density when the cavity is filled with helium. It is known that quasiparticles may travel long distances between superconducting islands on a substrate via conversion into phonons \cite{pat17}. It is possible that at intermediate temperatures, phonons in the superfluid helium may mediate the transfer of quasiparticles between the superconducting qubit and the superconducting \textit{cavity}. If the cavity were to have a higher nonequilibrium quasiparticle density, this additional coupling channel via the superfluid could cause the quasiparticle density in the qubit, and the associated relaxation rate via quasiparticle poisoning, to increase. At lower temperatures this additional source of quasiparticles would diminish as the Kapitza boundary resistance between the helium and the cavity/qubit continues to increase \cite{Pob92,Pol09}. This would lead to an increase in $T_1$ with decreasing temperature that would ultimately be limited by the same source that is limiting $T_1$ in the case of the empty cavity. 

Working within this hypothetical model, we partially fit, down to $\sim 60$ mK, the temperature dependent $T_1$ data for the case when the cavity is full of helium to the theoretical quasiparticle decay rate given in \cite{cat112} (see solid curve in Fig.~\ref{fig3}(a)). In this fit the only parameters are the normalized nonequilibrium quasiparticle density $x_{\mathrm{qp}}$ and the superconducting gap of aluminum $\Delta$. We find $\Delta \simeq160~\mu$eV, which agrees well with the known bulk value for aluminum, and that an increase in quasiparticle density of $\Delta x_{\mathrm{qp}} = 4\times10^{-6}$ accounts for the observed difference in $T_1$ above 60~mK when the cavity is filled with superfluid. 

If the helium is mediating the introduction of extra quasiparticles into the qubit at intermediate temperatures, this increased quasiparticle density should also cause a shift in the resonant frequency of the qubit \cite{cat112} relative to its zero temperature value. An increase in quasiparticle density of $\Delta x_{\mathrm{qp}}\sim4\times10^{-6}$ would produce a shift of $\Delta\omega_{01}\approx-14$~kHz in the qubit frequency. In Fig.~\ref{fig3}(b) we plot the resonant frequency of the qubit as a function of temperature 
and find that the qubit frequency does in fact shift down appreciably in this intermediate temperature regime when the cavity is filled with helium. While this data is consistent with our hypothesis of an increased quasiparticle density at intermediate temperatures, our current experiment cannot directly confirm this model of superfluid phonon mediated quasiparticle coupling between the qubit and the 3D cavity, however future experiments employing a 3D copper cavity may be able to do so.

Finally, while the temperature dependence of our $T_1$ data can be predominantly understood from the perspective of athermal quasiparticle poisoning, there likely are other mechanisms affecting the qubit energy relaxation. In particular the presence of near resonant two level system (TLS) defects \cite{mul15, bur19, sch19} can be a potential source of decreased $T_1$. We discuss the role of two level systems in the long timescale fluctuations we observe in $T_1$ in Appendix C.

\subsection{Qubit dephasing}

In contrast to the energy relaxation of the qubit, we find that above $60~$mK the pure dephasing time $T_{\phi}$ is the same both when the 3D cavity is empty and when it is full of superfluid. Upon cooling below $\sim60~$mK we find that the dephasing time modestly improves in the presence of helium, indicating that qubit dephasing and energy relaxation are dominated by different mechanisms. Experiments similar to ours are known to be plagued by thermal photon occupations well above the nominal temperature of the mixing chamber of the dilution refrigerator \cite{yan16,yan18,yeh17,wan19}, and fluctuations in the cavity photon number have been identified as a major limitation to the phase coherence of transmon qubits \cite{sea12,yan18,wan19}. Additionally, in these previous experiments (see particularly \cite{sea12,yeh17,yan18}) the temperature dependence of the dephasing rate is qualitatively similar to that which we observe in our measurements both with and without helium. In what follows we discuss our observation of an increased dephasing time when the cavity is filled with superfluid in the context of its possible effect on the thermal photon population of the 3D cavity and in Appendix C we discuss the role of other possible dephasing mechanisms.

It is known that many components in the microwave circuit are inefficiently thermalized, and it has been suggested that dissipative components such as attenuators may heat the microwave environment within the cavity to temperatures well above the dilution refrigerator temperature \cite{yeh17,wan19}. It is possible that the superfluid helium is serving to better thermalize the microwave environment within the cavity in our experiment. For example, the helium could be opening an additional channel to cool the microwave circuitry via the central pins of microwave coupling lines or by directly cooling the 3D cavity walls. The modest improvement we observe in qubit dephasing when our cavity is filled with superfluid would correspond to a relatively minor reduction in the thermal photon number in the cavity. For dephasing arising from residual thermal photons in the cavity, in the limit $\kappa \ll \chi$  where $\kappa$ is the cavity linewidth, and $\chi$ is the shift in the qubit frequency per cavity photon, we can express the dephasing rate as $\Gamma_{\phi}~=~\bar{n}_{th}\kappa\chi^2/(\kappa^2~+~\chi^2)$ \cite{cle07, wan19}, where $\bar{n}_{th}$ is the thermal population of photons in the cavity. Using this expression we can estimate the temperature of the photon bath $T_{ph}\sim80$ mK for the empty cavity and $T_{ph}\sim70$ mK for the superfluid filled cavity. Finally, we note that these results are reproducible over multiple cool-down cycles of the cryostat.

\subsection{Residual excited state population}

\begin{figure}
\begin{center}
\includegraphics[width=1 \columnwidth]{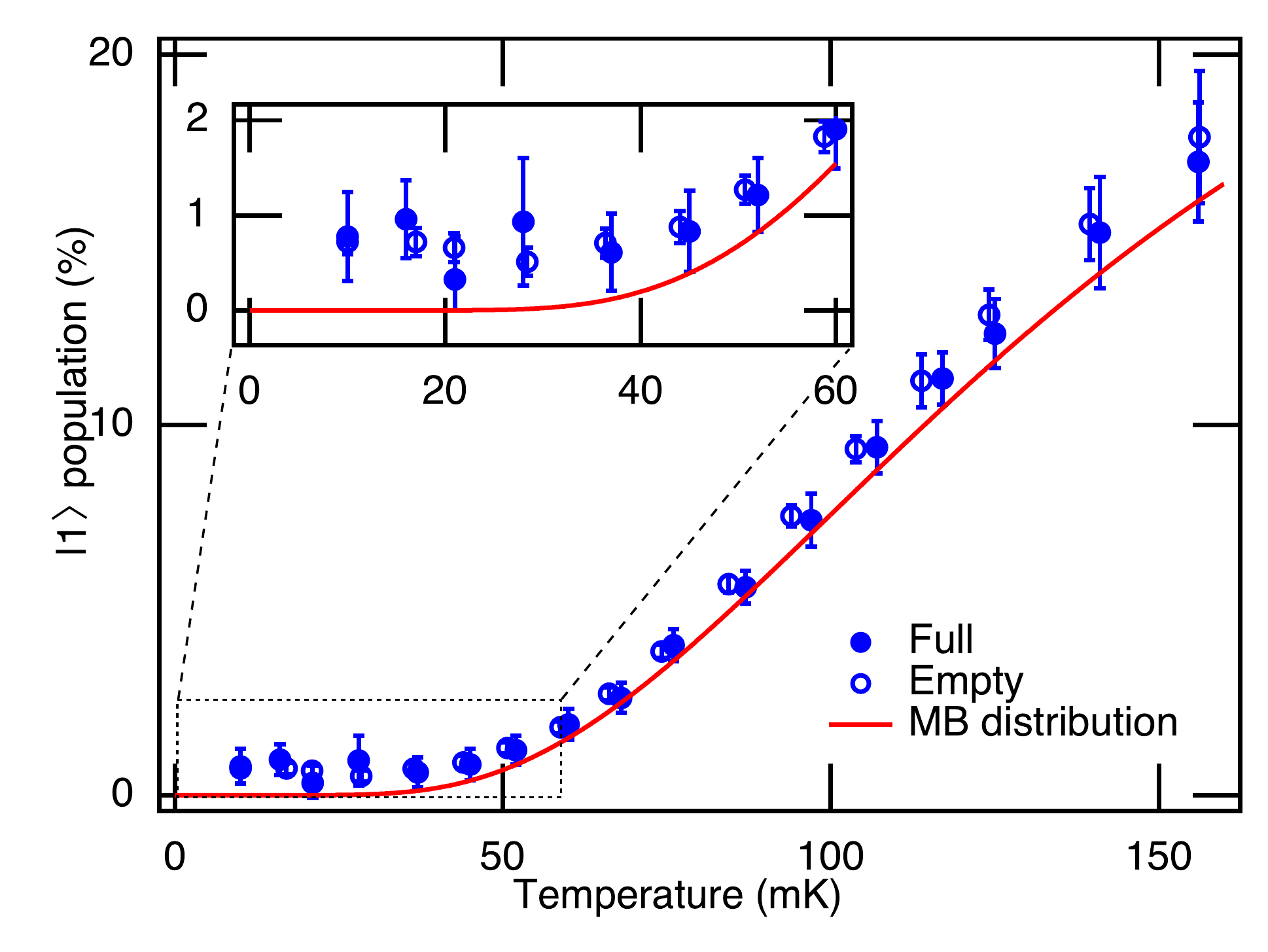}
\caption{(Color online) Residual population of the qubit excited state $\exc$ as a function of temperature for both the empty and superfluid filled cavity configurations. Also plotted is the theoretical Maxwell-Boltzman (MB) distribution, calculated using the energy levels obtained from spectroscopy. The presence of helium in the cavity has no significant effect on the $\exc$ population, and the data fit the expected population well with no adjustable parameters. Inset: expanded view of boxed region. The $\exc$ population for the qubit saturates at roughly the same value both when the cavity is full of superfluid and when it is empty.} 
\label{fig4}
\end{center}
\end{figure}

To further investigate possible thermalizing effects produced by the superfluid helium we have directly measured the residual qubit excited state population using a method developed in Refs. \cite{gee13,jin15}. The measured population is plotted as a function of temperature in Fig.~\ref{fig4}, along with the expected population calculated from a Maxwell-Boltzmann distribution with a partition function truncated beyond the $3^{\text{rd}}$ excited state of the qubit. As shown in Fig.~\ref{fig4}, the data are in good agreement with the theoretical population calculated with no adjustable parameters. Apparently, the superfluid helium has no significant effect on the residual excited state population of the qubit, which saturates at $0.5\%-1\%$ at the lowest temperature in both the empty and full cavity configurations. This is consistent with the known difficulty of effectively cooling $^{4}$He in the low milli-Kelvin temperature range due to the Kapitza thermal boundary resistance \cite{Pob92, del17} between superfluid helium and solid materials. Additionally, recent experiments attribute the majority of the residual excited state population to athermal quasiparticle poisioning \cite{ser18}. The saturation of the qubit excited state at roughly the same value independent of the presence of superfluid helium is therefore also consistent with our measurements of $T_1$ (which we find to also saturate at roughly the same value) being mainly limited by athermal quasiparticles at low temperatures. 

\section{Conclusion}
In conclusion, we have measured the effects of superfluid helium on the spectroscopic and coherence properties of a single-junction superconducting transmon qubit housed in an hermetically sealed 3D aluminum cavity that serves as an immersion cell. Within the framework of cQED, our experiments confirm that superfluid helium can be used to produce fine-tuning of the cavity frequency \textit{in situ} without significantly degrading its quality factor. We observe no seriously deleterious effects on the qubit coherence at the low operating temperatures of superconducting qubit experiments, and even modest improvement in qubit dephasing in the presence of the superfluid. These result show that quantum fluids are a potential tool for investigating coherence in ever increasingly higher quality superconducting qubit systems. In this vein, a possible extension of this work would be exploring the properties of a transmon qubit immersed in liquid $^3$He, which is known \cite{Pob92, Pol09} to thermalize to lower temperatures than $^4$He. Performing these experiments in a copper 3D cavity would further assist in the overall thermalization of the system and also allow for a frequency tunable qubit that could be used to study the local spectral distribution of two level systems coupled to the transmon (similar to Ref.~\cite{kli18}).


From a different perspective, combined superfluid/qubit systems could play a role in investigating fundamental aspects of quantum fluids. Since our results demonstrate that superfluid helium does not significantly degrade the quality of the qubit or the 3D cavity it is possible to imagine new experiments utilizing cQED to this end. Future devices could be modified to couple the qubit to specific mechanical modes of the superfluid contained in the cavity. For example, increasing the surface participation ratio of the qubit and covering it with a thin superfluid film via capillary action could allow for new qubit-assisted experiments to investigate excitations of the superfluid surface such as third sound \cite{Eve64, Sch98} or ripplons \cite{Saa75,dyk17}.

\begin{acknowledgements}
We are grateful to R. McDermott, D.I. Schuster, P.M. Harrington, M.I. Dykman, N.O. Birge, G. Koolstra, J. Kitzman and S. Hemmerle for fruitful discussions. We thank J.P. Davis and G.G. Popowich for technical assistance with proper superfluid thermalization. We also thank R. Loloee and B. Bi for technical assistance, and use of the W.M. Keck Microfabrication Facility at MSU. The Michigan State portion of this work was supported by the Cowen Family Endowment and by the NSF (Grant no. DMR-1708331). The Washington University portion of this work was also supported by the NSF (Grant no. PHY-1607156 and PHY-1752844 (CAREER)). 
\end{acknowledgements}

\appendix
\section{Experimental setup}
\begin{figure}[h!]
\centering
\includegraphics[width=8cm]{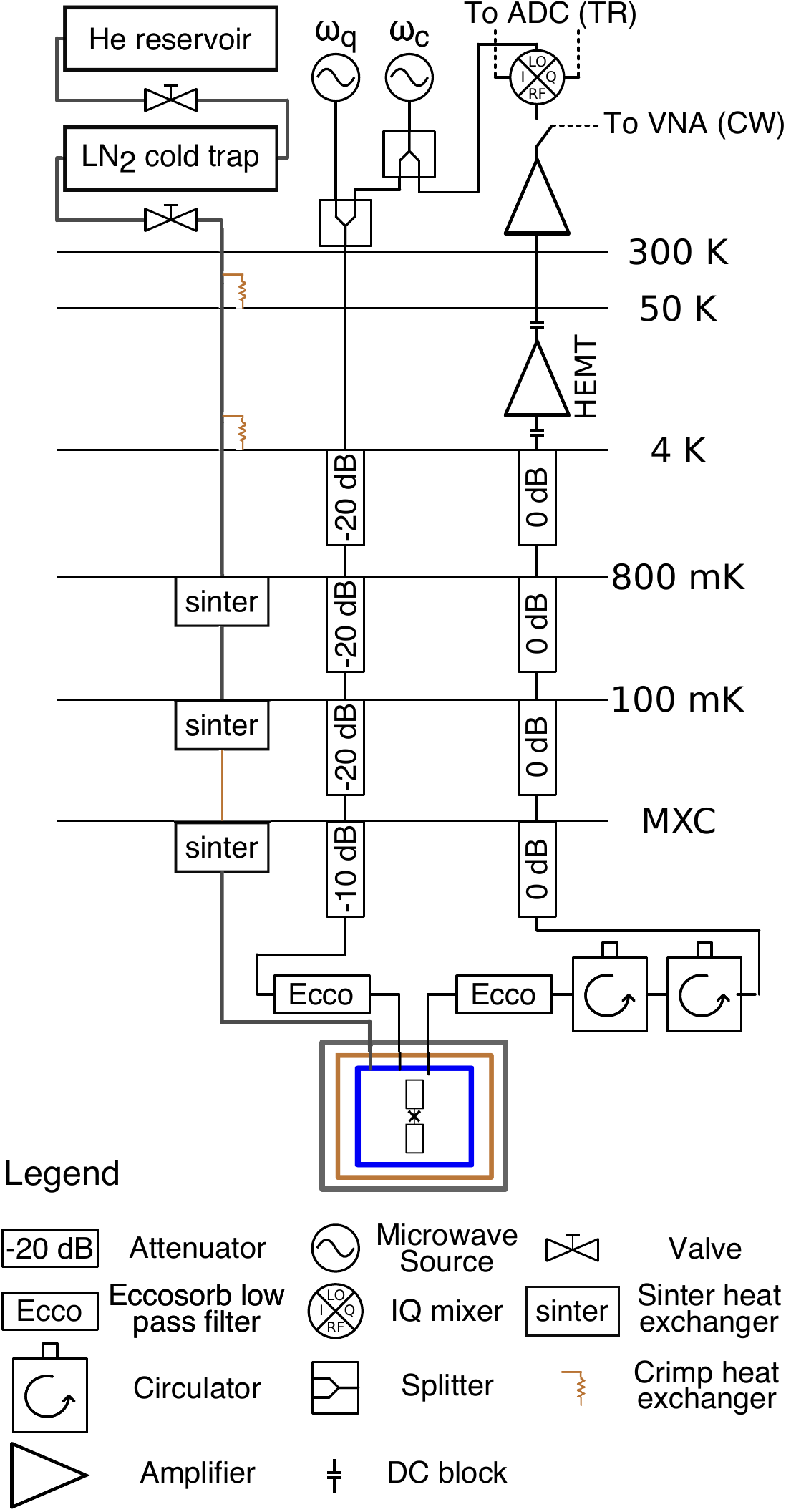}
\caption{(Color online) Schematic diagram of the experimental setup}
\end{figure}

The qubit is housed in a hermetically sealed 3D microwave cavity (blue box, Fig.~1 main text). Helium is introduced into the cavity from room temperature though a liquid nitrogen cooled cold trap to prevent impurities solidifying in the fill capillary and causing a blockage. The capillary line is a 1/16'' diameter stainless steel capillary (grey) everywhere except between the cold plate and mixing chamber plate, where a 0.017'' CuNi capillary (brown) is used to minimize heat flow between these two plates by a superfluid film within the capillary. The incoming helium is thermalized at five points: twice by mechanically clamping the fill capillary to the 50~K and 4~K plates of the cryostat, and by a passing the helium through a copper sinter heat exchanger at each of the still plate (800~mK), cold plate (100~mK), and mixing chamber plate (MXC).

The hermetically sealed 3D cavity (blue) is placed inside a light tight cooper box (brown) and cryogenic magnetic shielding (gray). The hermetic microwave assemblies are attached to a standard circuit cQED setup, consisting of distributed attenuators/circulators on the input/output lines respectively. For continuous wave (CW) measurements, the output signal is diverted to a vector network analyzer (VNA), while for time resolved (TR) measurements the output is demodulated by an IQ mixer and sent to an analog to digital converter (ADC) for acquisition.

\section{Dielectric dependence of the vacuum Rabi coupling}
\begin{figure}[h!]
\centering
\includegraphics[width=6cm]{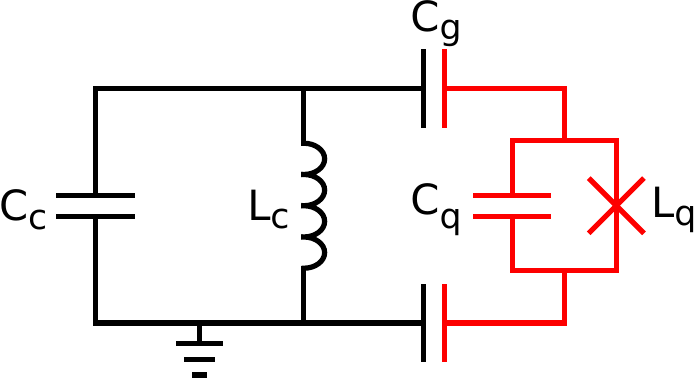}
\caption{(Color online) Circuit model for a 3D transmon circuit (red) coupled to a linear cavity (black).}
\end{figure}
We develop a simple model to capture the change of the vacuum Rabi coupling $g_{01}$ as a function of the dielectric constant of the environment surrounding the qubit. The vacuum Rabi coupling of $\gnd$ to $\exc$ may be written \cite{koc07} as 
\begin{equation}
g_{01} = 2eV_{ZPF} \beta \langle1\vert \hat{n} \vert 0 \rangle
\end{equation}
where $V_{ZPF}$ is the magnitude of the zero point fluctuations of the voltage in the cavity, $e$ is the electron charge, $\beta$ is a parameter that describes the efficiency with which voltage in the cavity builds up across the Josephson junction, and $\hat{n}$ is the Cooper pair excitation number operator. We model the cavity as a simple LC oscillator, which allows us to write the zero point fluctuations of the voltage in the cavity as \cite{dev97}
\begin{equation}
V_{ZPF} = \omega_c\sqrt{\frac{\hbar Z_c}{2}}
\end{equation}
where $\omega_c = 1/\sqrt{L_cC_c}$ is the resonant frequency and $Z_c = \sqrt{L_c/C_c}$ is the impedance of the oscillator. Uniformly filling the cavity with a dielectric will shift the cavity capacitance from $C_c$ to $\epsilon C_c$, and from Eq.~2 one finds that 
\begin{equation}
V_{ZPF} \propto \epsilon^{-3/4} 
\end{equation}

To understand the functional dependences of $\beta$ for our experiment, we model the qubit as a parallel capacitance \footnote{This capacitance includes both the intrinsic Josephson junction capacitance and the shunt capacitance provided by the antenna paddles of the qubit.} $C_q$ and nonlinear Josephson inductor $L_q$ coupled to the cavity via capacitance $C_g$ (see Fig.~2). We assume the system is symmetric and that both of the antenna paddles of the qubit are identical and have the same capacitance $C_g$ to the 3D cavity walls. $\beta$ is then given by the voltage that builds up across $C_q$ when a voltage $V$ exists across the entire circuit,

\begin{equation}
\beta = \frac{C_g^2}{C_g^2 + 2C_qC_g}
\end{equation}

\begin{figure*}[t]
\centering
\includegraphics[width=16cm]{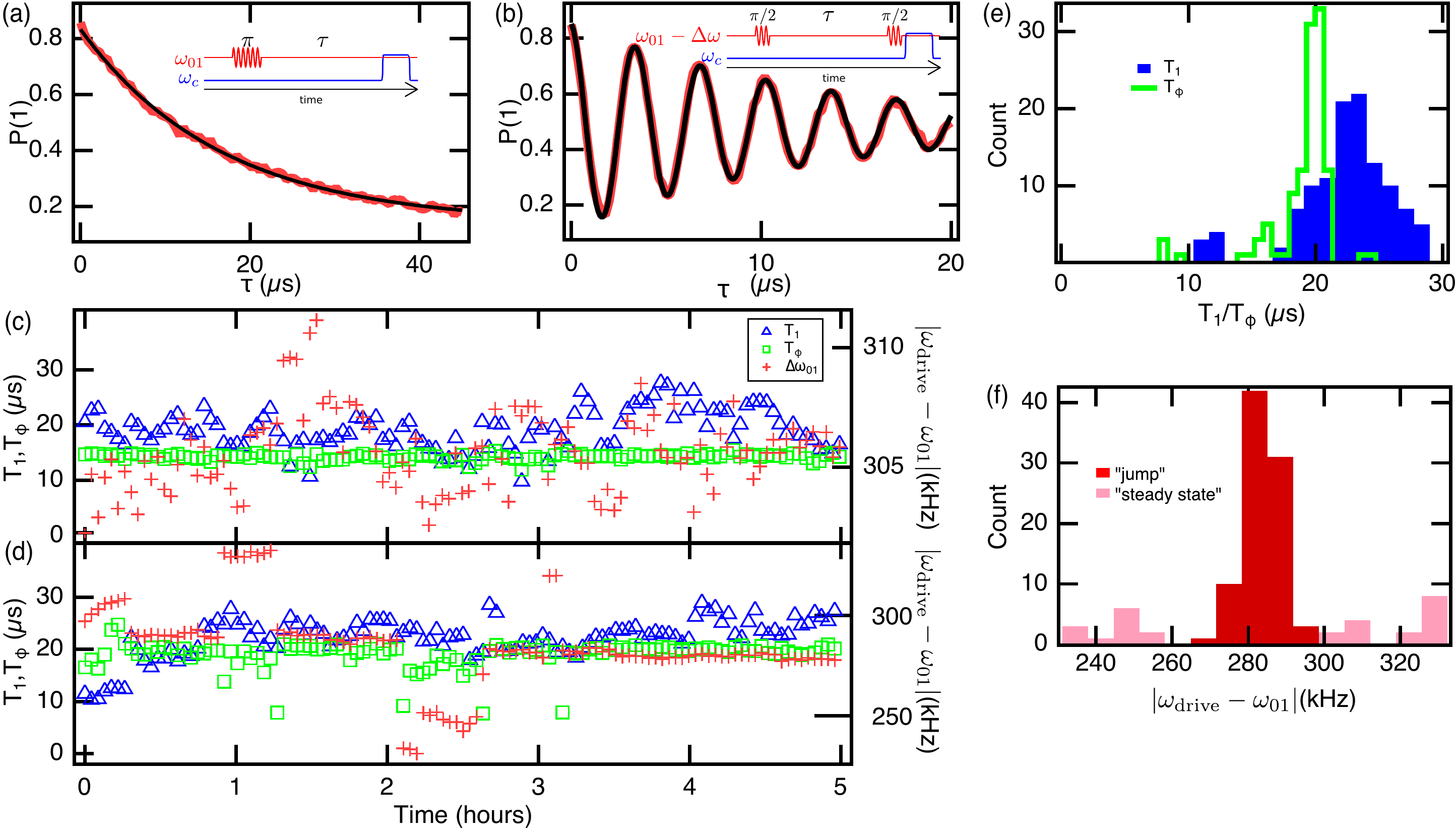}
\caption{(Color online) (a) Representative measurement of the qubit energy relaxation time $T_1$ inset with a schematic of the corresponding pulse sequence. We measure the probability $P(1)$ of finding the qubit in the excited state $\exc$ after a variable delay time $\tau$ after exciting it and fit the data to an exponential function to extract the decay time $T_1$. (b) A representative free induction decay measurement, which is fit to a sinusoid superimposed on a decaying exponential function. From this fit we extract the dephasing time $T_2$ and the drive/qubit detuning $\abs{\omega_{\mathrm{drive}} - \omega_{01}}$.  We interleave a single run of each measurement described in (a) and (b), and then repeat this for $\sim150$~s to get a single data set to fit to. We repeat this process for 5 hours to gather statistics on long timescale fluctuations: (c) is a representative measurement of $T_1, T_\phi= (1/T_2 - 1/2T_1)^{-1}$, and $\abs{\omega_{\mathrm{drive}} - \omega_{01}}$ for the empty cavity at $T\simeq10~$mK. (d) Is a similar measurement run at the same temperature but for the cavity filled with helium. Note the difference in scale between right axes of (c) and (d). (e) A histogram of the values of $T_1$ and $T_\phi$ plotted in (d): The average and standard deviations of these datasets are what is reported in Fig (\ref{fig3}) of the main text.  (f) A histogram of the values of $\Delta\omega_{01}$ recorded in (d): to measure the frequency shift as a function of temperature, we reject data measured during a discrete ``jump'' (light) and average over only data in ``steady state'' values of $\omega_{01}$ (dark). }
\label{fig7}
\end{figure*}

In the case of uniform dielectric filling, the capacitances will all scale uniformly and $\beta$ will be unchanged from its vacuum value. We note, however, that the presence of the silicon chip and the intrinsic Josephson junction capacitance will cause $C_g$ and $C_q$ to scale differently as a function of dielectric constant of the cavity medium. We perform finite element simulations using COMSOL to determine how these capacitances changes in the presence of helium. We find that $\Delta C_q = 0.78\%$ upon filling the cavity with helium, which agrees very well with the measured shift of $0.82\%$ extracted from the change in the charging energy $E_C$, and that $\Delta C_g = 1.65\%$.

We finally note that the transmon excitation number transition matrix element is proportional to the zero point charge fluctuations of the qubit, which in the nearly harmonic oscillator regime of the transmon circuit may be written as
\begin{equation}	
\abs{\bra{j+1}\hat{n}\ket{j}} \approx \sqrt{\frac{j+1}{2}} \bigg( \frac{E_J}{8E_C}\bigg)^{1/4}
\propto E_C^{-1/4} \propto C_q^{1/4}
\end{equation}
We use the simulated shift in $C_q$ to calculate the expected change of the charge number matrix element. Combining the shift in the matrix element with the predicted shifts in $V_{ZPF}$ and $\beta$, we arrive at a predicted shift in the vacuum Rabi coupling induced by the superfluid in the cavity
\begin{equation}
\Delta g_{01} = -3.2\%,
\end{equation}
which is in good agreement with our measured value of $-2.8\%$.

\section{Qubit coherence measurements}

To measure the qubit energy relaxation rate $T_1$, we use a standard measurement scheme, consisting of a $\pi$ pulse at the $\omega_{01}$ transition followed by a variable delay $\tau$ after which we projectively measure the state of the qubit \cite{ree10} (see inset of Fig (\ref{fig7}a)). We repeat this measurement while varying $\tau$, and fit the data to a decaying exponential function to extract $T_1$ (see Fig (\ref{fig7})). We use a similar free induction decay measurement to extract $T_2$, applying first a $\pi/2$ pulse detuned from $\omega_{01}$ by $\sim300$~kHz followed by variable delay $\tau$ before the application of a second $\pi/2$ pulse and measurement of the qubit state (inset of  Fig (\ref{fig7}b)). We fit the resulting data to a sinusoid superimposed on a decaying exponential function, and extract the qubit-drive detuning $\abs{\omega_{\mathrm{drive}}-\omega_{01}}$ (and subsequently $\omega_{01}$) from the frequency of the sinusoidal fit and $T_2$ from the exponentially decaying envelope. Additionally we conducted preliminary echo experiments for both the empty and superfluid-filled cavity configurations at the lowest temperature and found that the spin echo time was not significantly different than the Ramsey decay time.

To account for long time scale fluctuations of the qubit decoherence, we repeat the measurements of both $T_1$ and $T_2$ over a span of 5 hours. In Fig \ref{fig7}(c-d) we plot the energy relaxation time $T_1$, the pure dephasing time $T_\phi$ and $\abs{\omega_{\mathrm{drive}}-\omega_{01}}$ as a function of time for the empty (c) and superfluid filled (d) cavity configurations at the lowest temperature of the dilution refrigerator (10~mK). As the coherence times of superconducting qubits are routinely observed to fluctuate significantly over the span of several hours (see Fig \ref{fig7}(e-f)), in Fig \ref{fig3} of the main text we report the average and standard deviation of one such set of coherence measurements.

\subsection{Discussion of long timescale fluctuations in qubit properties}
We observe discrete abrupt changes (``jumps'') in the qubit frequency $\omega_{01}$ of the order $25-50$~kHz that occur over $\sim$hour timescales similar to those reported in other cQED experiments. We commonly observe (see Fig \ref{fig7}(c)) that $\omega_{01}$ will jump from some steady state value, stay at the new value for minutes to hours, and then return to the original steady state value. Discrete changes such as these are commonly attributed to critical current noise in the Josephson junction \cite{pai11}. Importantly, we can rule out shifts in the quasiparticle density as the origin of these jumps in $\omega_{01}$, as they would be accompanied by an associated shift in the qubit relaxation rate $\Gamma_1 = 1/T_1$ of order $25-12~\mu s^{-1}$ \cite{cat112}, which we do not observe. Therefore, to extract the temperature dependent quasiparticle induced shift in $\omega_{01}$ shown in Fig.~\ref{fig3}(b) of the main text, we employ a clustering algorithm to bin the measurements of $\omega_{01}$ around the steady state frequency and reject measurements that occur during a discrete change in the qubit frequency. Figure \ref{fig7}(f) shows an example of this for the data trace in Fig \ref{fig7}(d): the central (dark) data are accepted while the outlying (light) data that was recorded during a discrete jump in $\omega_{01}$ is rejected.

\begin{figure}[t!]
\centering
\includegraphics[width = 8cm]{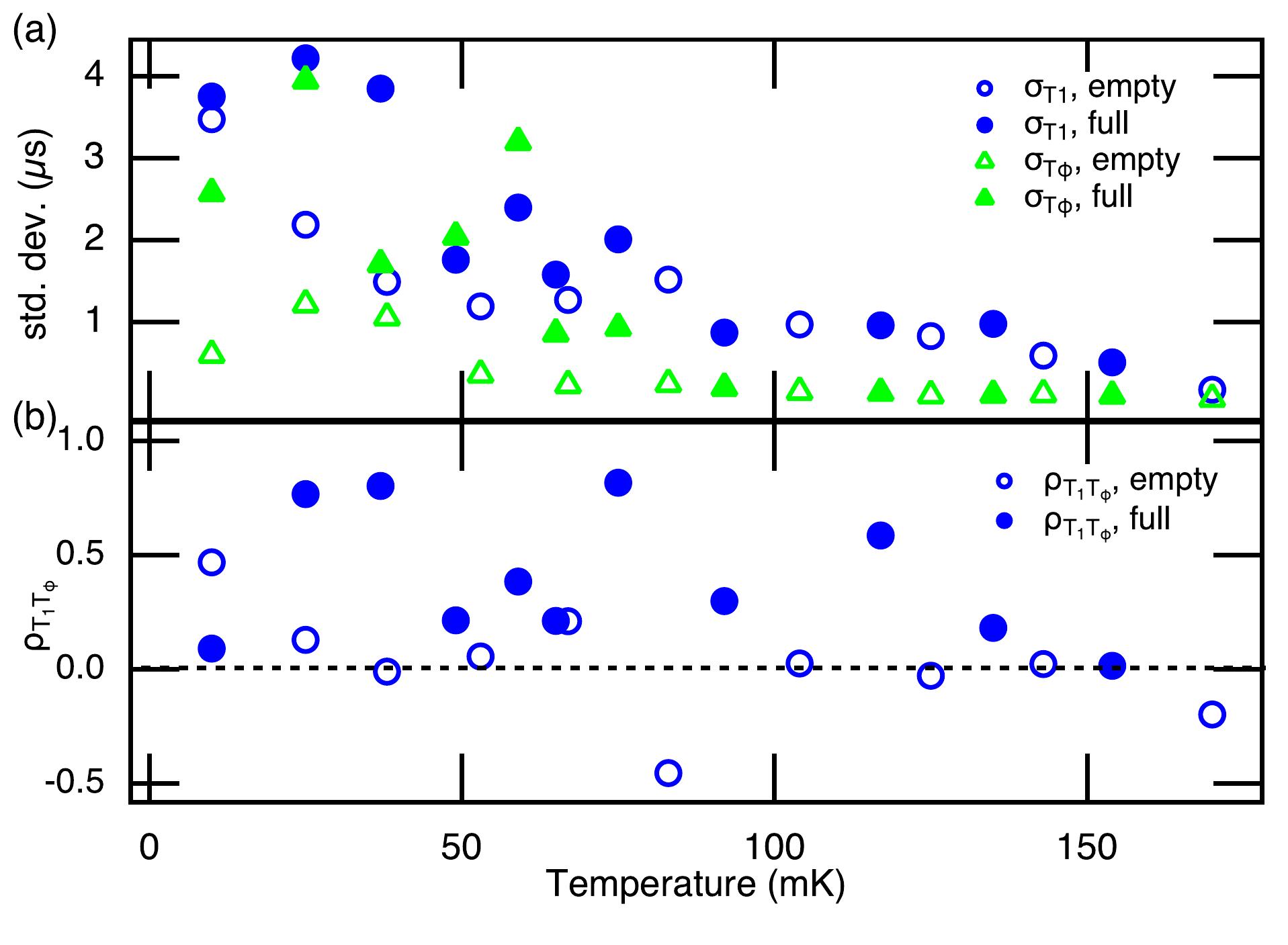}
\caption{(Color online) (a) Standard deviation of the measured datasets of $T_1$ and $T_\phi$ for both empty and full cavities. (b) Normalized covariance of the sets of measured $T_1$ and $T_\phi$.}
\label{fig8}
\end{figure}

In addition, we also observe significant long timescale fluctuations of the decay and decoherence times $T_1$ and $T_\phi$, which are not associated with changes we observe in $\omega_{01}$. Several recent studies \cite{mul15,bur19,sch19,kli18} have attributed the long timescale fluctuations in $T_1$ to two level systems (TLSs) in proximity to the qubit both spectrally and in real space. As these TLSs fluctuate in frequency, they potentially provide a time-varying density of states into which the qubit can lose energy. In Fig (\ref{fig8}) we plot the standard deviation $\sigma_{T1}~(\sigma_{T\phi})$ of each set of $T_1$ ($T_\phi$) measurements as a function of temperature and see that, generally, at lower temperatures the magnitude of the fluctuations is increased when the cavity is full of helium. These results indicate that unsaturated TLS fluctuators could be playing a role in qubit decoherence at the lowest temperatures of our experiment when the cavity is full of helium. However, further experiments optimized for spectral \cite{bur19,sch19} or time domain \cite{kli18} analysis of these fluctuations would be needed to confirm the roll that liquid helium has on TLS thermalization/fluctuation.

\subsection{Discussion of possible additional qubit dephasing mechanisms}
In addition to qubit dephasing produced by a fluctuating cavity photon number, several dephasing mechanisms are known to be important to superconducting qubits. Single junction transmon qubits, like ours, are known to be insensitive to both charge and flux noise \cite{sea12}, and the dephasing caused by quasiparticle poisoning is predicted to be negligible compared to the relaxation induced by quasiparticles (i.e. $\Gamma_{\phi,\mathrm{qp}} \ll \Gamma_{1,\mathrm{qp}}/2$ \cite{cat12}.) Another possible source of dephasing is from TLSs nearly resonant with the qubit: as these TLSs undergo spectral diffusion, the associated dispersive shifts will also fluctuate, leading to qubit dephasing. Fluctuations of near resonant TLSs can also create a fluctuating density of states into which the qubit can decay. If TLSs are a dominant source of both transverse and longitudinal noise we should expect some correlation between the fluctuations in $T_1$ and $T_\phi$. 

We plot the normalized covariance of $T_1$ and $T_\phi$, $\rho_{T_1,T_\phi} = (\langle T_1 T_\phi \rangle - \langle T_1 \rangle \langle T_\phi \rangle)/(\sigma_{T_1}\sigma_{T_\phi})$, where $\rho_{T_1,T_\phi} = 1$ corresponds to perfectly correlated values of $T_1$ and $T_\phi$ and $\rho_{T_1,T_\phi} = 0$ corresponds to $T_1$ and $T_\phi$ being completely uncorrelated. We find that when the cavity is empty there is little correlation between $T_1$ and $T_\phi$, and that this is broadly true also for the case when the cavity is filled with helium. There are, however, several measurement sets when the cavity is full of superfluid helium where $\rho_{T_1,T_\phi} \approx 0.8$, indicating strong correlation between $T_1$ and $T_\phi$ during these measurements. However, the more pronounced systematic increase in $T_\phi$ when the qubit is immersed in helium taken along with the fact that the correlation between $T_1$ and $T_\phi$ does not show any clear systematic temperature dependence indicates that, while perhaps not a dominant mechanism, TLSs could be playing a relatively larger role in the qubit energy decay and dephasing in the presence of superfluid helium for at least a subset of the measurements.

\bibliographystyle{apsrev4-1} 

\begin{thebibliography}{75}%
\makeatletter
\providecommand \@ifxundefined [1]{%
 \@ifx{#1\undefined}
}%
\providecommand \@ifnum [1]{%
 \ifnum #1\expandafter \@firstoftwo
 \else \expandafter \@secondoftwo
 \fi
}%
\providecommand \@ifx [1]{%
 \ifx #1\expandafter \@firstoftwo
 \else \expandafter \@secondoftwo
 \fi
}%
\providecommand \natexlab [1]{#1}%
\providecommand \enquote  [1]{``#1''}%
\providecommand \bibnamefont  [1]{#1}%
\providecommand \bibfnamefont [1]{#1}%
\providecommand \citenamefont [1]{#1}%
\providecommand \href@noop [0]{\@secondoftwo}%
\providecommand \href [0]{\begingroup \@sanitize@url \@href}%
\providecommand \@href[1]{\@@startlink{#1}\@@href}%
\providecommand \@@href[1]{\endgroup#1\@@endlink}%
\providecommand \@sanitize@url [0]{\catcode `\\12\catcode `\$12\catcode
  `\&12\catcode `\#12\catcode `\^12\catcode `\_12\catcode `\%12\relax}%
\providecommand \@@startlink[1]{}%
\providecommand \@@endlink[0]{}%
\providecommand \url  [0]{\begingroup\@sanitize@url \@url }%
\providecommand \@url [1]{\endgroup\@href {#1}{\urlprefix }}%
\providecommand \urlprefix  [0]{URL }%
\providecommand \Eprint [0]{\href }%
\providecommand \doibase [0]{http://dx.doi.org/}%
\providecommand \selectlanguage [0]{\@gobble}%
\providecommand \bibinfo  [0]{\@secondoftwo}%
\providecommand \bibfield  [0]{\@secondoftwo}%
\providecommand \translation [1]{[#1]}%
\providecommand \BibitemOpen [0]{}%
\providecommand \bibitemStop [0]{}%
\providecommand \bibitemNoStop [0]{.\EOS\space}%
\providecommand \EOS [0]{\spacefactor3000\relax}%
\providecommand \BibitemShut  [1]{\csname bibitem#1\endcsname}%
\let\auto@bib@innerbib\@empty
\bibitem [{\citenamefont {Wendin}(2017)}]{wen17}%
  \BibitemOpen
  \bibfield  {author} {\bibinfo {author} {\bibfnamefont {G.}~\bibnamefont
  {Wendin}},\ }\href@noop {} {\bibfield  {journal} {\bibinfo  {journal}
  {Reports on Progress in Physics}\ }\textbf {\bibinfo {volume} {80}},\
  \bibinfo {pages} {106001} (\bibinfo {year} {2017})}\BibitemShut {NoStop}%
\bibitem [{\citenamefont {Krinner}\ \emph {et~al.}(2019)\citenamefont
  {Krinner}, \citenamefont {Storz}, \citenamefont {Kurpiers}, \citenamefont
  {Magnard}, \citenamefont {Heinsoo}, \citenamefont {Keller}, \citenamefont
  {Luetolf}, \citenamefont {Eichler},\ and\ \citenamefont {Wallraff}}]{kri19}%
  \BibitemOpen
  \bibfield  {author} {\bibinfo {author} {\bibfnamefont {S.}~\bibnamefont
  {Krinner}}, \bibinfo {author} {\bibfnamefont {S.}~\bibnamefont {Storz}},
  \bibinfo {author} {\bibfnamefont {P.}~\bibnamefont {Kurpiers}}, \bibinfo
  {author} {\bibfnamefont {P.}~\bibnamefont {Magnard}}, \bibinfo {author}
  {\bibfnamefont {J.}~\bibnamefont {Heinsoo}}, \bibinfo {author} {\bibfnamefont
  {R.}~\bibnamefont {Keller}}, \bibinfo {author} {\bibfnamefont
  {J.}~\bibnamefont {Luetolf}}, \bibinfo {author} {\bibfnamefont
  {C.}~\bibnamefont {Eichler}}, \ and\ \bibinfo {author} {\bibfnamefont
  {A.}~\bibnamefont {Wallraff}},\ }\href@noop {} {\bibfield  {journal}
  {\bibinfo  {journal} {EPJ Quantum Technology}\ }\textbf {\bibinfo {volume}
  {6}},\ \bibinfo {pages} {2} (\bibinfo {year} {2019})}\BibitemShut {NoStop}%
\bibitem [{\citenamefont {Krantz}\ \emph {et~al.}(2019)\citenamefont {Krantz},
  \citenamefont {Kjaergaard}, \citenamefont {Yan}, \citenamefont {Orlando},
  \citenamefont {Gustavsson},\ and\ \citenamefont {Oliver}}]{kra19}%
  \BibitemOpen
  \bibfield  {author} {\bibinfo {author} {\bibfnamefont {P.}~\bibnamefont
  {Krantz}}, \bibinfo {author} {\bibfnamefont {M.}~\bibnamefont {Kjaergaard}},
  \bibinfo {author} {\bibfnamefont {F.}~\bibnamefont {Yan}}, \bibinfo {author}
  {\bibfnamefont {T.}~\bibnamefont {Orlando}}, \bibinfo {author} {\bibfnamefont
  {S.}~\bibnamefont {Gustavsson}}, \ and\ \bibinfo {author} {\bibfnamefont
  {W.}~\bibnamefont {Oliver}},\ }\href@noop {} {\bibfield  {journal} {\bibinfo
  {journal} {arXiv:190406560}\ } (\bibinfo {year} {2019})}\BibitemShut
  {NoStop}%
\bibitem [{\citenamefont {Wallraff}\ \emph {et~al.}(2004)\citenamefont
  {Wallraff}, \citenamefont {Schuster}, \citenamefont {Blais}, \citenamefont
  {Frunzio}, \citenamefont {Huang}, \citenamefont {Majer}, \citenamefont
  {Kumar}, \citenamefont {Girvin},\ and\ \citenamefont {Schoelkopf}}]{wal04}%
  \BibitemOpen
  \bibfield  {author} {\bibinfo {author} {\bibfnamefont {A.}~\bibnamefont
  {Wallraff}}, \bibinfo {author} {\bibfnamefont {D.~I.}\ \bibnamefont
  {Schuster}}, \bibinfo {author} {\bibfnamefont {A.}~\bibnamefont {Blais}},
  \bibinfo {author} {\bibfnamefont {L.}~\bibnamefont {Frunzio}}, \bibinfo
  {author} {\bibfnamefont {R.~S.}\ \bibnamefont {Huang}}, \bibinfo {author}
  {\bibfnamefont {J.}~\bibnamefont {Majer}}, \bibinfo {author} {\bibfnamefont
  {S.}~\bibnamefont {Kumar}}, \bibinfo {author} {\bibfnamefont {S.~M.}\
  \bibnamefont {Girvin}}, \ and\ \bibinfo {author} {\bibfnamefont {R.~J.}\
  \bibnamefont {Schoelkopf}},\ }\href {http://dx.doi.org/10.1038/nature02851}
  {\bibfield  {journal} {\bibinfo  {journal} {Nature}\ }\textbf {\bibinfo
  {volume} {431}},\ \bibinfo {pages} {162 EP } (\bibinfo {year}
  {2004})}\BibitemShut {NoStop}%
\bibitem [{\citenamefont {Schoelkopf}\ and\ \citenamefont
  {Girvin}(2008)}]{sch08}%
  \BibitemOpen
  \bibfield  {author} {\bibinfo {author} {\bibfnamefont {R.~J.}\ \bibnamefont
  {Schoelkopf}}\ and\ \bibinfo {author} {\bibfnamefont {S.~M.}\ \bibnamefont
  {Girvin}},\ }\href {http://dx.doi.org/10.1038/451664a} {\bibfield  {journal}
  {\bibinfo  {journal} {Nature}\ }\textbf {\bibinfo {volume} {451}},\ \bibinfo
  {pages} {664 EP } (\bibinfo {year} {2008})}\BibitemShut {NoStop}%
\bibitem [{\citenamefont {Paik}\ \emph {et~al.}(2011)\citenamefont {Paik},
  \citenamefont {Schuster}, \citenamefont {Bishop}, \citenamefont {Kirchmair},
  \citenamefont {Catelani}, \citenamefont {Sears}, \citenamefont {Johnson},
  \citenamefont {Reagor}, \citenamefont {Frunzio}, \citenamefont {Glazman},
  \citenamefont {Girvin}, \citenamefont {Devoret},\ and\ \citenamefont
  {Schoelkopf}}]{pai11}%
  \BibitemOpen
  \bibfield  {author} {\bibinfo {author} {\bibfnamefont {H.}~\bibnamefont
  {Paik}}, \bibinfo {author} {\bibfnamefont {D.~I.}\ \bibnamefont {Schuster}},
  \bibinfo {author} {\bibfnamefont {L.~S.}\ \bibnamefont {Bishop}}, \bibinfo
  {author} {\bibfnamefont {G.}~\bibnamefont {Kirchmair}}, \bibinfo {author}
  {\bibfnamefont {G.}~\bibnamefont {Catelani}}, \bibinfo {author}
  {\bibfnamefont {A.~P.}\ \bibnamefont {Sears}}, \bibinfo {author}
  {\bibfnamefont {B.~R.}\ \bibnamefont {Johnson}}, \bibinfo {author}
  {\bibfnamefont {M.~J.}\ \bibnamefont {Reagor}}, \bibinfo {author}
  {\bibfnamefont {L.}~\bibnamefont {Frunzio}}, \bibinfo {author} {\bibfnamefont
  {L.~I.}\ \bibnamefont {Glazman}}, \bibinfo {author} {\bibfnamefont {S.~M.}\
  \bibnamefont {Girvin}}, \bibinfo {author} {\bibfnamefont {M.~H.}\
  \bibnamefont {Devoret}}, \ and\ \bibinfo {author} {\bibfnamefont {R.~J.}\
  \bibnamefont {Schoelkopf}},\ }\href {\doibase 10.1103/PhysRevLett.107.240501}
  {\bibfield  {journal} {\bibinfo  {journal} {Phys. Rev. Lett.}\ }\textbf
  {\bibinfo {volume} {107}},\ \bibinfo {pages} {240501} (\bibinfo {year}
  {2011})}\BibitemShut {NoStop}%
\bibitem [{\citenamefont {Rigetti}\ \emph {et~al.}(2012)\citenamefont
  {Rigetti}, \citenamefont {Gambetta}, \citenamefont {Poletto}, \citenamefont
  {Plourde}, \citenamefont {Chow}, \citenamefont {C\'orcoles}, \citenamefont
  {Smolin}, \citenamefont {Merkel}, \citenamefont {Rozen}, \citenamefont
  {Keefe}, \citenamefont {Rothwell}, \citenamefont {Ketchen},\ and\
  \citenamefont {Steffen}}]{rig12}%
  \BibitemOpen
  \bibfield  {author} {\bibinfo {author} {\bibfnamefont {C.}~\bibnamefont
  {Rigetti}}, \bibinfo {author} {\bibfnamefont {J.~M.}\ \bibnamefont
  {Gambetta}}, \bibinfo {author} {\bibfnamefont {S.}~\bibnamefont {Poletto}},
  \bibinfo {author} {\bibfnamefont {B.~L.~T.}\ \bibnamefont {Plourde}},
  \bibinfo {author} {\bibfnamefont {J.~M.}\ \bibnamefont {Chow}}, \bibinfo
  {author} {\bibfnamefont {A.~D.}\ \bibnamefont {C\'orcoles}}, \bibinfo
  {author} {\bibfnamefont {J.~A.}\ \bibnamefont {Smolin}}, \bibinfo {author}
  {\bibfnamefont {S.~T.}\ \bibnamefont {Merkel}}, \bibinfo {author}
  {\bibfnamefont {J.~R.}\ \bibnamefont {Rozen}}, \bibinfo {author}
  {\bibfnamefont {G.~A.}\ \bibnamefont {Keefe}}, \bibinfo {author}
  {\bibfnamefont {M.~B.}\ \bibnamefont {Rothwell}}, \bibinfo {author}
  {\bibfnamefont {M.~B.}\ \bibnamefont {Ketchen}}, \ and\ \bibinfo {author}
  {\bibfnamefont {M.}~\bibnamefont {Steffen}},\ }\href {\doibase
  10.1103/PhysRevB.86.100506} {\bibfield  {journal} {\bibinfo  {journal} {Phys.
  Rev. B}\ }\textbf {\bibinfo {volume} {86}},\ \bibinfo {pages} {100506}
  (\bibinfo {year} {2012})}\BibitemShut {NoStop}%
\bibitem [{\citenamefont {Reagor}\ \emph {et~al.}(2013)\citenamefont {Reagor},
  \citenamefont {Paik}, \citenamefont {Catelani}, \citenamefont {Sun},
  \citenamefont {Axline}, \citenamefont {Holland}, \citenamefont {Pop},
  \citenamefont {Masluk}, \citenamefont {Brecht}, \citenamefont {Frunzio},
  \citenamefont {Devoret}, \citenamefont {Glazman},\ and\ \citenamefont
  {Schoelkopf}}]{rea13}%
  \BibitemOpen
  \bibfield  {author} {\bibinfo {author} {\bibfnamefont {M.}~\bibnamefont
  {Reagor}}, \bibinfo {author} {\bibfnamefont {H.}~\bibnamefont {Paik}},
  \bibinfo {author} {\bibfnamefont {G.}~\bibnamefont {Catelani}}, \bibinfo
  {author} {\bibfnamefont {L.}~\bibnamefont {Sun}}, \bibinfo {author}
  {\bibfnamefont {C.}~\bibnamefont {Axline}}, \bibinfo {author} {\bibfnamefont
  {E.}~\bibnamefont {Holland}}, \bibinfo {author} {\bibfnamefont {I.~M.}\
  \bibnamefont {Pop}}, \bibinfo {author} {\bibfnamefont {N.~A.}\ \bibnamefont
  {Masluk}}, \bibinfo {author} {\bibfnamefont {T.}~\bibnamefont {Brecht}},
  \bibinfo {author} {\bibfnamefont {L.}~\bibnamefont {Frunzio}}, \bibinfo
  {author} {\bibfnamefont {M.~H.}\ \bibnamefont {Devoret}}, \bibinfo {author}
  {\bibfnamefont {L.}~\bibnamefont {Glazman}}, \ and\ \bibinfo {author}
  {\bibfnamefont {R.~J.}\ \bibnamefont {Schoelkopf}},\ }\href {\doibase
  10.1063/1.4807015} {\bibfield  {journal} {\bibinfo  {journal} {Applied
  Physics Letters}\ }\textbf {\bibinfo {volume} {102}},\ \bibinfo {pages}
  {192604} (\bibinfo {year} {2013})},\ \Eprint
  {http://arxiv.org/abs/https://doi.org/10.1063/1.4807015}
  {https://doi.org/10.1063/1.4807015} \BibitemShut {NoStop}%
\bibitem [{\citenamefont {Reagor}\ \emph {et~al.}(2016)\citenamefont {Reagor},
  \citenamefont {Pfaff}, \citenamefont {Axline}, \citenamefont {Heeres},
  \citenamefont {Ofek}, \citenamefont {Sliwa}, \citenamefont {Holland},
  \citenamefont {Wang}, \citenamefont {Blumoff}, \citenamefont {Chou},
  \citenamefont {Hatridge}, \citenamefont {Frunzio}, \citenamefont {Devoret},
  \citenamefont {Jiang},\ and\ \citenamefont {Schoelkopf}}]{rea16}%
  \BibitemOpen
  \bibfield  {author} {\bibinfo {author} {\bibfnamefont {M.}~\bibnamefont
  {Reagor}}, \bibinfo {author} {\bibfnamefont {W.}~\bibnamefont {Pfaff}},
  \bibinfo {author} {\bibfnamefont {C.}~\bibnamefont {Axline}}, \bibinfo
  {author} {\bibfnamefont {R.~W.}\ \bibnamefont {Heeres}}, \bibinfo {author}
  {\bibfnamefont {N.}~\bibnamefont {Ofek}}, \bibinfo {author} {\bibfnamefont
  {K.}~\bibnamefont {Sliwa}}, \bibinfo {author} {\bibfnamefont
  {E.}~\bibnamefont {Holland}}, \bibinfo {author} {\bibfnamefont
  {C.}~\bibnamefont {Wang}}, \bibinfo {author} {\bibfnamefont {J.}~\bibnamefont
  {Blumoff}}, \bibinfo {author} {\bibfnamefont {K.}~\bibnamefont {Chou}},
  \bibinfo {author} {\bibfnamefont {M.~J.}\ \bibnamefont {Hatridge}}, \bibinfo
  {author} {\bibfnamefont {L.}~\bibnamefont {Frunzio}}, \bibinfo {author}
  {\bibfnamefont {M.~H.}\ \bibnamefont {Devoret}}, \bibinfo {author}
  {\bibfnamefont {L.}~\bibnamefont {Jiang}}, \ and\ \bibinfo {author}
  {\bibfnamefont {R.~J.}\ \bibnamefont {Schoelkopf}},\ }\href {\doibase
  10.1103/PhysRevB.94.014506} {\bibfield  {journal} {\bibinfo  {journal} {Phys.
  Rev. B}\ }\textbf {\bibinfo {volume} {94}},\ \bibinfo {pages} {014506}
  (\bibinfo {year} {2016})}\BibitemShut {NoStop}%
\bibitem [{\citenamefont {Ofek}\ \emph {et~al.}(2016)\citenamefont {Ofek},
  \citenamefont {Petrenko}, \citenamefont {Heeres}, \citenamefont {Reinhold},
  \citenamefont {Leghtas}, \citenamefont {Vlastakis}, \citenamefont {Liu},
  \citenamefont {Frunzio}, \citenamefont {Girvin}, \citenamefont {Jiang},
  \citenamefont {Mirrahimi}, \citenamefont {Devoret},\ and\ \citenamefont
  {Schoelkopf}}]{ofe16}%
  \BibitemOpen
  \bibfield  {author} {\bibinfo {author} {\bibfnamefont {N.}~\bibnamefont
  {Ofek}}, \bibinfo {author} {\bibfnamefont {A.}~\bibnamefont {Petrenko}},
  \bibinfo {author} {\bibfnamefont {R.}~\bibnamefont {Heeres}}, \bibinfo
  {author} {\bibfnamefont {P.}~\bibnamefont {Reinhold}}, \bibinfo {author}
  {\bibfnamefont {Z.}~\bibnamefont {Leghtas}}, \bibinfo {author} {\bibfnamefont
  {B.}~\bibnamefont {Vlastakis}}, \bibinfo {author} {\bibfnamefont
  {Y.}~\bibnamefont {Liu}}, \bibinfo {author} {\bibfnamefont {L.}~\bibnamefont
  {Frunzio}}, \bibinfo {author} {\bibfnamefont {S.~M.}\ \bibnamefont {Girvin}},
  \bibinfo {author} {\bibfnamefont {L.}~\bibnamefont {Jiang}}, \bibinfo
  {author} {\bibfnamefont {M.}~\bibnamefont {Mirrahimi}}, \bibinfo {author}
  {\bibfnamefont {M.~H.}\ \bibnamefont {Devoret}}, \ and\ \bibinfo {author}
  {\bibfnamefont {R.~J.}\ \bibnamefont {Schoelkopf}},\ }\href
  {http://dx.doi.org/10.1038/nature18949} {\bibfield  {journal} {\bibinfo
  {journal} {Nature}\ }\textbf {\bibinfo {volume} {536}},\ \bibinfo {pages}
  {441 EP } (\bibinfo {year} {2016})}\BibitemShut {NoStop}%
\bibitem [{\citenamefont {Wenner}\ \emph {et~al.}(2011)\citenamefont {Wenner},
  \citenamefont {Neeley}, \citenamefont {Bialczak}, \citenamefont {Lenander},
  \citenamefont {Lucero}, \citenamefont {O'Connell}, \citenamefont {Sank},
  \citenamefont {Wang}, \citenamefont {Weides}, \citenamefont {Cleland},\ and\
  \citenamefont {Martinis}}]{wen11}%
  \BibitemOpen
  \bibfield  {author} {\bibinfo {author} {\bibfnamefont {J.}~\bibnamefont
  {Wenner}}, \bibinfo {author} {\bibfnamefont {M.}~\bibnamefont {Neeley}},
  \bibinfo {author} {\bibfnamefont {R.~C.}\ \bibnamefont {Bialczak}}, \bibinfo
  {author} {\bibfnamefont {M.}~\bibnamefont {Lenander}}, \bibinfo {author}
  {\bibfnamefont {E.}~\bibnamefont {Lucero}}, \bibinfo {author} {\bibfnamefont
  {A.~D.}\ \bibnamefont {O'Connell}}, \bibinfo {author} {\bibfnamefont
  {D.}~\bibnamefont {Sank}}, \bibinfo {author} {\bibfnamefont {H.}~\bibnamefont
  {Wang}}, \bibinfo {author} {\bibfnamefont {M.}~\bibnamefont {Weides}},
  \bibinfo {author} {\bibfnamefont {A.~N.}\ \bibnamefont {Cleland}}, \ and\
  \bibinfo {author} {\bibfnamefont {J.~M.}\ \bibnamefont {Martinis}},\ }\href
  {http://stacks.iop.org/0953-2048/24/i=6/a=065001} {\bibfield  {journal}
  {\bibinfo  {journal} {Superconductor Science and Technology}\ }\textbf
  {\bibinfo {volume} {24}},\ \bibinfo {pages} {065001} (\bibinfo {year}
  {2011})}\BibitemShut {NoStop}%
\bibitem [{\citenamefont {Devoret}\ and\ \citenamefont
  {Schoelkopf}(2013)}]{dev13}%
  \BibitemOpen
  \bibfield  {author} {\bibinfo {author} {\bibfnamefont {M.~H.}\ \bibnamefont
  {Devoret}}\ and\ \bibinfo {author} {\bibfnamefont {R.~J.}\ \bibnamefont
  {Schoelkopf}},\ }\href {\doibase 10.1126/science.1231930} {\bibfield
  {journal} {\bibinfo  {journal} {Science}\ }\textbf {\bibinfo {volume}
  {339}},\ \bibinfo {pages} {1169} (\bibinfo {year} {2013})},\ \Eprint
  {http://arxiv.org/abs/http://science.sciencemag.org/content/339/6124/1169.full.pdf}
  {http://science.sciencemag.org/content/339/6124/1169.full.pdf} \BibitemShut
  {NoStop}%
\bibitem [{\citenamefont {Brecht}\ \emph {et~al.}(2016)\citenamefont {Brecht},
  \citenamefont {Pfaff}, \citenamefont {Wang}, \citenamefont {Chu},
  \citenamefont {Frunzio}, \citenamefont {Devoret},\ and\ \citenamefont
  {Schoelkopf}}]{bre16}%
  \BibitemOpen
  \bibfield  {author} {\bibinfo {author} {\bibfnamefont {T.}~\bibnamefont
  {Brecht}}, \bibinfo {author} {\bibfnamefont {W.}~\bibnamefont {Pfaff}},
  \bibinfo {author} {\bibfnamefont {C.}~\bibnamefont {Wang}}, \bibinfo {author}
  {\bibfnamefont {Y.}~\bibnamefont {Chu}}, \bibinfo {author} {\bibfnamefont
  {L.}~\bibnamefont {Frunzio}}, \bibinfo {author} {\bibfnamefont {M.~H.}\
  \bibnamefont {Devoret}}, \ and\ \bibinfo {author} {\bibfnamefont {R.~J.}\
  \bibnamefont {Schoelkopf}},\ }\href {https://doi.org/10.1038/npjqi.2016.2}
  {\bibfield  {journal} {\bibinfo  {journal} {Npj Quantum Information}\
  }\textbf {\bibinfo {volume} {2}},\ \bibinfo {pages} {16002 EP } (\bibinfo
  {year} {2016})}\BibitemShut {NoStop}%
\bibitem [{Note1()}]{Note1}%
  \BibitemOpen
  \bibinfo {note} {We note that \protect \emph {two-dimensional} resonators can
  be frequency tuned, albeit with a reduced quality factor, via the
  incorporation of a SQUID loop in series with the resonator center conductor
  \cite {Pal08, San08}}\BibitemShut {NoStop}%
\bibitem [{\citenamefont {Yeh}\ \emph {et~al.}(2017)\citenamefont {Yeh},
  \citenamefont {LeFebvre}, \citenamefont {Premaratne}, \citenamefont
  {Wellstood},\ and\ \citenamefont {Palmer}}]{yeh17}%
  \BibitemOpen
  \bibfield  {author} {\bibinfo {author} {\bibfnamefont {J.-H.}\ \bibnamefont
  {Yeh}}, \bibinfo {author} {\bibfnamefont {J.}~\bibnamefont {LeFebvre}},
  \bibinfo {author} {\bibfnamefont {S.}~\bibnamefont {Premaratne}}, \bibinfo
  {author} {\bibfnamefont {F.~C.}\ \bibnamefont {Wellstood}}, \ and\ \bibinfo
  {author} {\bibfnamefont {B.~S.}\ \bibnamefont {Palmer}},\ }\href {\doibase
  10.1063/1.4984894} {\bibfield  {journal} {\bibinfo  {journal} {Journal of
  Applied Physics}\ }\textbf {\bibinfo {volume} {121}},\ \bibinfo {pages}
  {224501} (\bibinfo {year} {2017})},\ \Eprint
  {http://arxiv.org/abs/https://doi.org/10.1063/1.4984894}
  {https://doi.org/10.1063/1.4984894} \BibitemShut {NoStop}%
\bibitem [{\citenamefont {Yan}\ \emph {et~al.}(2018)\citenamefont {Yan},
  \citenamefont {Campbell}, \citenamefont {Krantz}, \citenamefont {Kjaergaard},
  \citenamefont {Kim}, \citenamefont {Yoder}, \citenamefont {Hover},
  \citenamefont {Sears}, \citenamefont {Kerman}, \citenamefont {Orlando},
  \citenamefont {Gustavsson},\ and\ \citenamefont {Oliver}}]{yan18}%
  \BibitemOpen
  \bibfield  {author} {\bibinfo {author} {\bibfnamefont {F.}~\bibnamefont
  {Yan}}, \bibinfo {author} {\bibfnamefont {D.}~\bibnamefont {Campbell}},
  \bibinfo {author} {\bibfnamefont {P.}~\bibnamefont {Krantz}}, \bibinfo
  {author} {\bibfnamefont {M.}~\bibnamefont {Kjaergaard}}, \bibinfo {author}
  {\bibfnamefont {D.}~\bibnamefont {Kim}}, \bibinfo {author} {\bibfnamefont
  {J.~L.}\ \bibnamefont {Yoder}}, \bibinfo {author} {\bibfnamefont
  {D.}~\bibnamefont {Hover}}, \bibinfo {author} {\bibfnamefont
  {A.}~\bibnamefont {Sears}}, \bibinfo {author} {\bibfnamefont {A.~J.}\
  \bibnamefont {Kerman}}, \bibinfo {author} {\bibfnamefont {T.~P.}\
  \bibnamefont {Orlando}}, \bibinfo {author} {\bibfnamefont {S.}~\bibnamefont
  {Gustavsson}}, \ and\ \bibinfo {author} {\bibfnamefont {W.~D.}\ \bibnamefont
  {Oliver}},\ }\href {\doibase 10.1103/PhysRevLett.120.260504} {\bibfield
  {journal} {\bibinfo  {journal} {Phys. Rev. Lett.}\ }\textbf {\bibinfo
  {volume} {120}},\ \bibinfo {pages} {260504} (\bibinfo {year}
  {2018})}\BibitemShut {NoStop}%
\bibitem [{\citenamefont {Wang}\ \emph {et~al.}(2019)\citenamefont {Wang},
  \citenamefont {Shankar}, \citenamefont {Minev}, \citenamefont
  {Campagne-Ibarcq}, \citenamefont {Narla},\ and\ \citenamefont
  {Devoret}}]{wan19}%
  \BibitemOpen
  \bibfield  {author} {\bibinfo {author} {\bibfnamefont {Z.}~\bibnamefont
  {Wang}}, \bibinfo {author} {\bibfnamefont {S.}~\bibnamefont {Shankar}},
  \bibinfo {author} {\bibfnamefont {Z.}~\bibnamefont {Minev}}, \bibinfo
  {author} {\bibfnamefont {P.}~\bibnamefont {Campagne-Ibarcq}}, \bibinfo
  {author} {\bibfnamefont {A.}~\bibnamefont {Narla}}, \ and\ \bibinfo {author}
  {\bibfnamefont {M.}~\bibnamefont {Devoret}},\ }\href {\doibase
  10.1103/PhysRevApplied.11.014031} {\bibfield  {journal} {\bibinfo  {journal}
  {Phys. Rev. Applied}\ }\textbf {\bibinfo {volume} {11}},\ \bibinfo {pages}
  {014031} (\bibinfo {year} {2019})}\BibitemShut {NoStop}%
\bibitem [{\citenamefont {Martinis}\ \emph {et~al.}(2009)\citenamefont
  {Martinis}, \citenamefont {Ansmann},\ and\ \citenamefont
  {Aumentado}}]{mar09}%
  \BibitemOpen
  \bibfield  {author} {\bibinfo {author} {\bibfnamefont {J.~M.}\ \bibnamefont
  {Martinis}}, \bibinfo {author} {\bibfnamefont {M.}~\bibnamefont {Ansmann}}, \
  and\ \bibinfo {author} {\bibfnamefont {J.}~\bibnamefont {Aumentado}},\ }\href
  {\doibase 10.1103/PhysRevLett.103.097002} {\bibfield  {journal} {\bibinfo
  {journal} {Phys. Rev. Lett.}\ }\textbf {\bibinfo {volume} {103}},\ \bibinfo
  {pages} {097002} (\bibinfo {year} {2009})}\BibitemShut {NoStop}%
\bibitem [{\citenamefont {Catelani}\ \emph {et~al.}(2011)\citenamefont
  {Catelani}, \citenamefont {Schoelkopf}, \citenamefont {Devoret},\ and\
  \citenamefont {Glazman}}]{cat112}%
  \BibitemOpen
  \bibfield  {author} {\bibinfo {author} {\bibfnamefont {G.}~\bibnamefont
  {Catelani}}, \bibinfo {author} {\bibfnamefont {R.~J.}\ \bibnamefont
  {Schoelkopf}}, \bibinfo {author} {\bibfnamefont {M.~H.}\ \bibnamefont
  {Devoret}}, \ and\ \bibinfo {author} {\bibfnamefont {L.~I.}\ \bibnamefont
  {Glazman}},\ }\href {\doibase 10.1103/PhysRevB.84.064517} {\bibfield
  {journal} {\bibinfo  {journal} {Phys. Rev. B}\ }\textbf {\bibinfo {volume}
  {84}},\ \bibinfo {pages} {064517} (\bibinfo {year} {2011})}\BibitemShut
  {NoStop}%
\bibitem [{\citenamefont {Patel}\ \emph {et~al.}(2017)\citenamefont {Patel},
  \citenamefont {Pechenezhskiy}, \citenamefont {Plourde}, \citenamefont
  {Vavilov},\ and\ \citenamefont {McDermott}}]{pat17}%
  \BibitemOpen
  \bibfield  {author} {\bibinfo {author} {\bibfnamefont {U.}~\bibnamefont
  {Patel}}, \bibinfo {author} {\bibfnamefont {I.~V.}\ \bibnamefont
  {Pechenezhskiy}}, \bibinfo {author} {\bibfnamefont {B.~L.~T.}\ \bibnamefont
  {Plourde}}, \bibinfo {author} {\bibfnamefont {M.~G.}\ \bibnamefont
  {Vavilov}}, \ and\ \bibinfo {author} {\bibfnamefont {R.}~\bibnamefont
  {McDermott}},\ }\href {\doibase 10.1103/PhysRevB.96.220501} {\bibfield
  {journal} {\bibinfo  {journal} {Phys. Rev. B}\ }\textbf {\bibinfo {volume}
  {96}},\ \bibinfo {pages} {220501} (\bibinfo {year} {2017})}\BibitemShut
  {NoStop}%
\bibitem [{\citenamefont {Serniak}\ \emph {et~al.}(2018)\citenamefont
  {Serniak}, \citenamefont {Hays}, \citenamefont {de~Lange}, \citenamefont
  {Diamond}, \citenamefont {Shankar}, \citenamefont {Burkhart}, \citenamefont
  {Frunzio}, \citenamefont {Houzet},\ and\ \citenamefont {Devoret}}]{ser18}%
  \BibitemOpen
  \bibfield  {author} {\bibinfo {author} {\bibfnamefont {K.}~\bibnamefont
  {Serniak}}, \bibinfo {author} {\bibfnamefont {M.}~\bibnamefont {Hays}},
  \bibinfo {author} {\bibfnamefont {G.}~\bibnamefont {de~Lange}}, \bibinfo
  {author} {\bibfnamefont {S.}~\bibnamefont {Diamond}}, \bibinfo {author}
  {\bibfnamefont {S.}~\bibnamefont {Shankar}}, \bibinfo {author} {\bibfnamefont
  {L.~D.}\ \bibnamefont {Burkhart}}, \bibinfo {author} {\bibfnamefont
  {L.}~\bibnamefont {Frunzio}}, \bibinfo {author} {\bibfnamefont
  {M.}~\bibnamefont {Houzet}}, \ and\ \bibinfo {author} {\bibfnamefont {M.~H.}\
  \bibnamefont {Devoret}},\ }\href {\doibase 10.1103/PhysRevLett.121.157701}
  {\bibfield  {journal} {\bibinfo  {journal} {Phys. Rev. Lett.}\ }\textbf
  {\bibinfo {volume} {121}},\ \bibinfo {pages} {157701} (\bibinfo {year}
  {2018})}\BibitemShut {NoStop}%
\bibitem [{\citenamefont {Samkharadze}\ \emph {et~al.}(2011)\citenamefont
  {Samkharadze}, \citenamefont {Kumar}, \citenamefont {Manfra}, \citenamefont
  {Pfeiffer}, \citenamefont {West},\ and\ \citenamefont {Cs{\'a}thy}}]{sam11}%
  \BibitemOpen
  \bibfield  {author} {\bibinfo {author} {\bibfnamefont {N.}~\bibnamefont
  {Samkharadze}}, \bibinfo {author} {\bibfnamefont {A.}~\bibnamefont {Kumar}},
  \bibinfo {author} {\bibfnamefont {M.~J.}\ \bibnamefont {Manfra}}, \bibinfo
  {author} {\bibfnamefont {L.~N.}\ \bibnamefont {Pfeiffer}}, \bibinfo {author}
  {\bibfnamefont {K.~W.}\ \bibnamefont {West}}, \ and\ \bibinfo {author}
  {\bibfnamefont {G.~A.}\ \bibnamefont {Cs{\'a}thy}},\ }\href {\doibase
  10.1063/1.3586766} {\bibfield  {journal} {\bibinfo  {journal} {Review of
  Scientific Instruments}\ }\textbf {\bibinfo {volume} {82}},\ \bibinfo {pages}
  {053902} (\bibinfo {year} {2011})},\ \Eprint
  {http://arxiv.org/abs/https://aip.scitation.org/doi/pdf/10.1063/1.3586766}
  {https://aip.scitation.org/doi/pdf/10.1063/1.3586766} \BibitemShut {NoStop}%
\bibitem [{\citenamefont {Bradley}\ \emph {et~al.}(2016)\citenamefont
  {Bradley}, \citenamefont {George}, \citenamefont {Gunnarsson}, \citenamefont
  {Haley}, \citenamefont {Heikkinen}, \citenamefont {Pashkin}, \citenamefont
  {Penttil{\"a}}, \citenamefont {Prance}, \citenamefont {Prunnila},
  \citenamefont {Roschier},\ and\ \citenamefont {Sarsby}}]{bra16}%
  \BibitemOpen
  \bibfield  {author} {\bibinfo {author} {\bibfnamefont {D.~I.}\ \bibnamefont
  {Bradley}}, \bibinfo {author} {\bibfnamefont {R.~E.}\ \bibnamefont {George}},
  \bibinfo {author} {\bibfnamefont {D.}~\bibnamefont {Gunnarsson}}, \bibinfo
  {author} {\bibfnamefont {R.~P.}\ \bibnamefont {Haley}}, \bibinfo {author}
  {\bibfnamefont {H.}~\bibnamefont {Heikkinen}}, \bibinfo {author}
  {\bibfnamefont {Y.~A.}\ \bibnamefont {Pashkin}}, \bibinfo {author}
  {\bibfnamefont {J.}~\bibnamefont {Penttil{\"a}}}, \bibinfo {author}
  {\bibfnamefont {J.~R.}\ \bibnamefont {Prance}}, \bibinfo {author}
  {\bibfnamefont {M.}~\bibnamefont {Prunnila}}, \bibinfo {author}
  {\bibfnamefont {L.}~\bibnamefont {Roschier}}, \ and\ \bibinfo {author}
  {\bibfnamefont {M.}~\bibnamefont {Sarsby}},\ }\href {\doibase
  10.1038/ncomms10455} {\bibfield  {journal} {\bibinfo  {journal} {Nature
  Communications}\ }\textbf {\bibinfo {volume} {7}},\ \bibinfo {pages} {10455}
  (\bibinfo {year} {2016})}\BibitemShut {NoStop}%
\bibitem [{\citenamefont {Souris}\ \emph {et~al.}(2017)\citenamefont {Souris},
  \citenamefont {Christiani},\ and\ \citenamefont {Davis}}]{sou17}%
  \BibitemOpen
  \bibfield  {author} {\bibinfo {author} {\bibfnamefont {F.}~\bibnamefont
  {Souris}}, \bibinfo {author} {\bibfnamefont {H.}~\bibnamefont {Christiani}},
  \ and\ \bibinfo {author} {\bibfnamefont {J.~P.}\ \bibnamefont {Davis}},\
  }\href {\doibase 10.1063/1.4997641} {\bibfield  {journal} {\bibinfo
  {journal} {Applied Physics Letters}\ }\textbf {\bibinfo {volume} {111}},\
  \bibinfo {pages} {172601} (\bibinfo {year} {2017})},\ \Eprint
  {http://arxiv.org/abs/https://doi.org/10.1063/1.4997641}
  {https://doi.org/10.1063/1.4997641} \BibitemShut {NoStop}%
\bibitem [{\citenamefont {Lorenzo}\ and\ \citenamefont {Schwab}(2014)}]{del14}%
  \BibitemOpen
  \bibfield  {author} {\bibinfo {author} {\bibfnamefont {L.~A.~D.}\
  \bibnamefont {Lorenzo}}\ and\ \bibinfo {author} {\bibfnamefont {K.~C.}\
  \bibnamefont {Schwab}},\ }\href {\doibase 10.1088/1367-2630/16/11/113020}
  {\bibfield  {journal} {\bibinfo  {journal} {New Journal of Physics}\ }\textbf
  {\bibinfo {volume} {16}},\ \bibinfo {pages} {113020} (\bibinfo {year}
  {2014})}\BibitemShut {NoStop}%
\bibitem [{\citenamefont {De~Lorenzo}\ and\ \citenamefont
  {Schwab}(2017)}]{del17}%
  \BibitemOpen
  \bibfield  {author} {\bibinfo {author} {\bibfnamefont {L.~A.}\ \bibnamefont
  {De~Lorenzo}}\ and\ \bibinfo {author} {\bibfnamefont {K.~C.}\ \bibnamefont
  {Schwab}},\ }\href {\doibase 10.1007/s10909-016-1674-x} {\bibfield  {journal}
  {\bibinfo  {journal} {Journal of Low Temperature Physics}\ }\textbf {\bibinfo
  {volume} {186}},\ \bibinfo {pages} {233} (\bibinfo {year}
  {2017})}\BibitemShut {NoStop}%
\bibitem [{Note2()}]{Note2}%
  \BibitemOpen
  \bibinfo {note} {Additionally, recent experiments have demonstrated that
  superfluid helium can be used as the working fluid in a mechanically actuated
  3D microwave cavity having a tunability $> 5$ GHz \cite {cla18}}\BibitemShut
  {NoStop}%
\bibitem [{\citenamefont {Harris}\ \emph {et~al.}(2016)\citenamefont {Harris},
  \citenamefont {McAuslan}, \citenamefont {Sheridan}, \citenamefont {Sachkou},
  \citenamefont {Baker},\ and\ \citenamefont {Bowen}}]{har16}%
  \BibitemOpen
  \bibfield  {author} {\bibinfo {author} {\bibfnamefont {G.~I.}\ \bibnamefont
  {Harris}}, \bibinfo {author} {\bibfnamefont {D.~L.}\ \bibnamefont
  {McAuslan}}, \bibinfo {author} {\bibfnamefont {E.}~\bibnamefont {Sheridan}},
  \bibinfo {author} {\bibfnamefont {Y.}~\bibnamefont {Sachkou}}, \bibinfo
  {author} {\bibfnamefont {C.}~\bibnamefont {Baker}}, \ and\ \bibinfo {author}
  {\bibfnamefont {W.~P.}\ \bibnamefont {Bowen}},\ }\href
  {https://doi.org/10.1038/nphys3714} {\bibfield  {journal} {\bibinfo
  {journal} {Nature Physics}\ }\textbf {\bibinfo {volume} {12}},\ \bibinfo
  {pages} {788 EP } (\bibinfo {year} {2016})}\BibitemShut {NoStop}%
\bibitem [{\citenamefont {Childress}\ \emph {et~al.}(2017)\citenamefont
  {Childress}, \citenamefont {Schmidt}, \citenamefont {Kashkanova},
  \citenamefont {Brown}, \citenamefont {Harris}, \citenamefont {Aiello},
  \citenamefont {Marquardt},\ and\ \citenamefont {Harris}}]{chi17}%
  \BibitemOpen
  \bibfield  {author} {\bibinfo {author} {\bibfnamefont {L.}~\bibnamefont
  {Childress}}, \bibinfo {author} {\bibfnamefont {M.~P.}\ \bibnamefont
  {Schmidt}}, \bibinfo {author} {\bibfnamefont {A.~D.}\ \bibnamefont
  {Kashkanova}}, \bibinfo {author} {\bibfnamefont {C.~D.}\ \bibnamefont
  {Brown}}, \bibinfo {author} {\bibfnamefont {G.~I.}\ \bibnamefont {Harris}},
  \bibinfo {author} {\bibfnamefont {A.}~\bibnamefont {Aiello}}, \bibinfo
  {author} {\bibfnamefont {F.}~\bibnamefont {Marquardt}}, \ and\ \bibinfo
  {author} {\bibfnamefont {J.~G.~E.}\ \bibnamefont {Harris}},\ }\href {\doibase
  10.1103/PhysRevA.96.063842} {\bibfield  {journal} {\bibinfo  {journal} {Phys.
  Rev. A}\ }\textbf {\bibinfo {volume} {96}},\ \bibinfo {pages} {063842}
  (\bibinfo {year} {2017})}\BibitemShut {NoStop}%
\bibitem [{\citenamefont {Platzman}\ and\ \citenamefont
  {Dykman}(1999)}]{pla99}%
  \BibitemOpen
  \bibfield  {author} {\bibinfo {author} {\bibfnamefont {P.}~\bibnamefont
  {Platzman}}\ and\ \bibinfo {author} {\bibfnamefont {M.}~\bibnamefont
  {Dykman}},\ }\href@noop {} {\bibfield  {journal} {\bibinfo  {journal}
  {Science}\ }\textbf {\bibinfo {volume} {284}},\ \bibinfo {pages} {1967}
  (\bibinfo {year} {1999})}\BibitemShut {NoStop}%
\bibitem [{\citenamefont {Dykman}\ \emph {et~al.}(2003)\citenamefont {Dykman},
  \citenamefont {Platzman},\ and\ \citenamefont {Seddighrad}}]{dyk03}%
  \BibitemOpen
  \bibfield  {author} {\bibinfo {author} {\bibfnamefont {M.~I.}\ \bibnamefont
  {Dykman}}, \bibinfo {author} {\bibfnamefont {P.~M.}\ \bibnamefont
  {Platzman}}, \ and\ \bibinfo {author} {\bibfnamefont {P.}~\bibnamefont
  {Seddighrad}},\ }\href {\doibase 10.1103/PhysRevB.67.155402} {\bibfield
  {journal} {\bibinfo  {journal} {Phys. Rev. B}\ }\textbf {\bibinfo {volume}
  {67}},\ \bibinfo {pages} {155402} (\bibinfo {year} {2003})}\BibitemShut
  {NoStop}%
\bibitem [{\citenamefont {Lyon}(2006)}]{lyo06}%
  \BibitemOpen
  \bibfield  {author} {\bibinfo {author} {\bibfnamefont {S.}~\bibnamefont
  {Lyon}},\ }\href@noop {} {\bibfield  {journal} {\bibinfo  {journal} {Physical
  Review A}\ }\textbf {\bibinfo {volume} {74}},\ \bibinfo {pages} {052338}
  (\bibinfo {year} {2006})}\BibitemShut {NoStop}%
\bibitem [{\citenamefont {Schuster}\ \emph {et~al.}(2010)\citenamefont
  {Schuster}, \citenamefont {Fragner}, \citenamefont {Dykman}, \citenamefont
  {Lyon},\ and\ \citenamefont {Schoelkopf}}]{schu10}%
  \BibitemOpen
  \bibfield  {author} {\bibinfo {author} {\bibfnamefont {D.~I.}\ \bibnamefont
  {Schuster}}, \bibinfo {author} {\bibfnamefont {A.}~\bibnamefont {Fragner}},
  \bibinfo {author} {\bibfnamefont {M.~I.}\ \bibnamefont {Dykman}}, \bibinfo
  {author} {\bibfnamefont {S.~A.}\ \bibnamefont {Lyon}}, \ and\ \bibinfo
  {author} {\bibfnamefont {R.~J.}\ \bibnamefont {Schoelkopf}},\ }\href
  {\doibase 10.1103/PhysRevLett.105.040503} {\bibfield  {journal} {\bibinfo
  {journal} {Phys. Rev. Lett.}\ }\textbf {\bibinfo {volume} {105}},\ \bibinfo
  {pages} {040503} (\bibinfo {year} {2010})}\BibitemShut {NoStop}%
\bibitem [{\citenamefont {Yang}\ \emph {et~al.}(2016)\citenamefont {Yang},
  \citenamefont {Fragner}, \citenamefont {Koolstra}, \citenamefont {Ocola},
  \citenamefont {Czaplewski}, \citenamefont {Schoelkopf},\ and\ \citenamefont
  {Schuster}}]{yan162}%
  \BibitemOpen
  \bibfield  {author} {\bibinfo {author} {\bibfnamefont {G.}~\bibnamefont
  {Yang}}, \bibinfo {author} {\bibfnamefont {A.}~\bibnamefont {Fragner}},
  \bibinfo {author} {\bibfnamefont {G.}~\bibnamefont {Koolstra}}, \bibinfo
  {author} {\bibfnamefont {L.}~\bibnamefont {Ocola}}, \bibinfo {author}
  {\bibfnamefont {D.~A.}\ \bibnamefont {Czaplewski}}, \bibinfo {author}
  {\bibfnamefont {R.~J.}\ \bibnamefont {Schoelkopf}}, \ and\ \bibinfo {author}
  {\bibfnamefont {D.~I.}\ \bibnamefont {Schuster}},\ }\href {\doibase
  10.1103/PhysRevX.6.011031} {\bibfield  {journal} {\bibinfo  {journal} {Phys.
  Rev. X}\ }\textbf {\bibinfo {volume} {6}},\ \bibinfo {pages} {011031}
  (\bibinfo {year} {2016})}\BibitemShut {NoStop}%
\bibitem [{\citenamefont {Nasyedkin}\ \emph {et~al.}(2018)\citenamefont
  {Nasyedkin}, \citenamefont {Byeon}, \citenamefont {Zhang}, \citenamefont
  {Beysengulov}, \citenamefont {Milem}, \citenamefont {Hemmerle}, \citenamefont
  {Loloee},\ and\ \citenamefont {Pollanen}}]{nas18}%
  \BibitemOpen
  \bibfield  {author} {\bibinfo {author} {\bibfnamefont {K.}~\bibnamefont
  {Nasyedkin}}, \bibinfo {author} {\bibfnamefont {H.}~\bibnamefont {Byeon}},
  \bibinfo {author} {\bibfnamefont {L.}~\bibnamefont {Zhang}}, \bibinfo
  {author} {\bibfnamefont {N.}~\bibnamefont {Beysengulov}}, \bibinfo {author}
  {\bibfnamefont {J.}~\bibnamefont {Milem}}, \bibinfo {author} {\bibfnamefont
  {S.}~\bibnamefont {Hemmerle}}, \bibinfo {author} {\bibfnamefont
  {R.}~\bibnamefont {Loloee}}, \ and\ \bibinfo {author} {\bibfnamefont
  {J.}~\bibnamefont {Pollanen}},\ }\href@noop {} {\bibfield  {journal}
  {\bibinfo  {journal} {Journal of Physics: Condensed Matter}\ }\textbf
  {\bibinfo {volume} {30}} (\bibinfo {year} {2018})}\BibitemShut {NoStop}%
\bibitem [{\citenamefont {Koolstra}\ \emph {et~al.}(2019)\citenamefont
  {Koolstra}, \citenamefont {Yang},\ and\ \citenamefont {Schuster}}]{koo19}%
  \BibitemOpen
  \bibfield  {author} {\bibinfo {author} {\bibfnamefont {G.}~\bibnamefont
  {Koolstra}}, \bibinfo {author} {\bibfnamefont {G.}~\bibnamefont {Yang}}, \
  and\ \bibinfo {author} {\bibfnamefont {D.~I.}\ \bibnamefont {Schuster}},\
  }\href@noop {} {\bibfield  {journal} {\bibinfo  {journal} {arXiv:1902.04190}\
  } (\bibinfo {year} {2019})}\BibitemShut {NoStop}%
\bibitem [{\citenamefont {Byeon}\ \emph {et~al.}(2019)\citenamefont {Byeon},
  \citenamefont {Nasyedkin}, \citenamefont {Lane}, \citenamefont {Zhang},
  \citenamefont {Beysengulov}, \citenamefont {Loloee},\ and\ \citenamefont
  {Pollanen}}]{bye19}%
  \BibitemOpen
  \bibfield  {author} {\bibinfo {author} {\bibfnamefont {H.}~\bibnamefont
  {Byeon}}, \bibinfo {author} {\bibfnamefont {K.}~\bibnamefont {Nasyedkin}},
  \bibinfo {author} {\bibfnamefont {J.}~\bibnamefont {Lane}}, \bibinfo {author}
  {\bibfnamefont {L.}~\bibnamefont {Zhang}}, \bibinfo {author} {\bibfnamefont
  {N.}~\bibnamefont {Beysengulov}}, \bibinfo {author} {\bibfnamefont
  {R.}~\bibnamefont {Loloee}}, \ and\ \bibinfo {author} {\bibfnamefont
  {J.}~\bibnamefont {Pollanen}},\ }\href@noop {} {\bibfield  {journal}
  {\bibinfo  {journal} {Journal of Low Temperature Physics}\ }\textbf {\bibinfo
  {volume} {195}} (\bibinfo {year} {2019})}\BibitemShut {NoStop}%
\bibitem [{\citenamefont {O'Connell}\ \emph {et~al.}(2010)\citenamefont
  {O'Connell}, \citenamefont {Hofheinz}, \citenamefont {Ansmann}, \citenamefont
  {Bialczak}, \citenamefont {Lenander}, \citenamefont {Lucero}, \citenamefont
  {Neeley}, \citenamefont {Sank}, \citenamefont {Wang}, \citenamefont {Weides},
  \citenamefont {Wenner}, \citenamefont {Martinis},\ and\ \citenamefont
  {Cleland}}]{ocon10}%
  \BibitemOpen
  \bibfield  {author} {\bibinfo {author} {\bibfnamefont {A.~D.}\ \bibnamefont
  {O'Connell}}, \bibinfo {author} {\bibfnamefont {M.}~\bibnamefont {Hofheinz}},
  \bibinfo {author} {\bibfnamefont {M.}~\bibnamefont {Ansmann}}, \bibinfo
  {author} {\bibfnamefont {R.~C.}\ \bibnamefont {Bialczak}}, \bibinfo {author}
  {\bibfnamefont {M.}~\bibnamefont {Lenander}}, \bibinfo {author}
  {\bibfnamefont {E.}~\bibnamefont {Lucero}}, \bibinfo {author} {\bibfnamefont
  {M.}~\bibnamefont {Neeley}}, \bibinfo {author} {\bibfnamefont
  {D.}~\bibnamefont {Sank}}, \bibinfo {author} {\bibfnamefont {H.}~\bibnamefont
  {Wang}}, \bibinfo {author} {\bibfnamefont {M.}~\bibnamefont {Weides}},
  \bibinfo {author} {\bibfnamefont {J.}~\bibnamefont {Wenner}}, \bibinfo
  {author} {\bibfnamefont {J.~M.}\ \bibnamefont {Martinis}}, \ and\ \bibinfo
  {author} {\bibfnamefont {A.~N.}\ \bibnamefont {Cleland}},\ }\href
  {http://dx.doi.org/10.1038/nature08967} {\bibfield  {journal} {\bibinfo
  {journal} {Nature}\ }\textbf {\bibinfo {volume} {464}},\ \bibinfo {pages}
  {697 EP } (\bibinfo {year} {2010})}\BibitemShut {NoStop}%
\bibitem [{\citenamefont {Moores}\ \emph {et~al.}(2018)\citenamefont {Moores},
  \citenamefont {Sletten}, \citenamefont {Viennot},\ and\ \citenamefont
  {Lehnert}}]{moo18}%
  \BibitemOpen
  \bibfield  {author} {\bibinfo {author} {\bibfnamefont {B.~A.}\ \bibnamefont
  {Moores}}, \bibinfo {author} {\bibfnamefont {L.~R.}\ \bibnamefont {Sletten}},
  \bibinfo {author} {\bibfnamefont {J.~J.}\ \bibnamefont {Viennot}}, \ and\
  \bibinfo {author} {\bibfnamefont {K.~W.}\ \bibnamefont {Lehnert}},\ }\href
  {\doibase 10.1103/PhysRevLett.120.227701} {\bibfield  {journal} {\bibinfo
  {journal} {Phys. Rev. Lett.}\ }\textbf {\bibinfo {volume} {120}},\ \bibinfo
  {pages} {227701} (\bibinfo {year} {2018})}\BibitemShut {NoStop}%
\bibitem [{\citenamefont {Satzinger}\ \emph {et~al.}(2018)\citenamefont
  {Satzinger}, \citenamefont {Zhong}, \citenamefont {Chang}, \citenamefont
  {Peairs}, \citenamefont {Bienfait}, \citenamefont {Chou}, \citenamefont
  {Cleland}, \citenamefont {Conner}, \citenamefont {Dumur}, \citenamefont
  {Grebel}, \citenamefont {Gutierrez}, \citenamefont {November}, \citenamefont
  {Povey}, \citenamefont {Whiteley}, \citenamefont {Awschalom}, \citenamefont
  {Schuster},\ and\ \citenamefont {Cleland}}]{sat18}%
  \BibitemOpen
  \bibfield  {author} {\bibinfo {author} {\bibfnamefont {K.~J.}\ \bibnamefont
  {Satzinger}}, \bibinfo {author} {\bibfnamefont {Y.~P.}\ \bibnamefont
  {Zhong}}, \bibinfo {author} {\bibfnamefont {H.~S.}\ \bibnamefont {Chang}},
  \bibinfo {author} {\bibfnamefont {G.~A.}\ \bibnamefont {Peairs}}, \bibinfo
  {author} {\bibfnamefont {A.}~\bibnamefont {Bienfait}}, \bibinfo {author}
  {\bibfnamefont {M.-H.}\ \bibnamefont {Chou}}, \bibinfo {author}
  {\bibfnamefont {A.~Y.}\ \bibnamefont {Cleland}}, \bibinfo {author}
  {\bibfnamefont {C.~R.}\ \bibnamefont {Conner}}, \bibinfo {author}
  {\bibfnamefont {{\'E}.}~\bibnamefont {Dumur}}, \bibinfo {author}
  {\bibfnamefont {J.}~\bibnamefont {Grebel}}, \bibinfo {author} {\bibfnamefont
  {I.}~\bibnamefont {Gutierrez}}, \bibinfo {author} {\bibfnamefont {B.~H.}\
  \bibnamefont {November}}, \bibinfo {author} {\bibfnamefont {R.~G.}\
  \bibnamefont {Povey}}, \bibinfo {author} {\bibfnamefont {S.~J.}\ \bibnamefont
  {Whiteley}}, \bibinfo {author} {\bibfnamefont {D.~D.}\ \bibnamefont
  {Awschalom}}, \bibinfo {author} {\bibfnamefont {D.~I.}\ \bibnamefont
  {Schuster}}, \ and\ \bibinfo {author} {\bibfnamefont {A.~N.}\ \bibnamefont
  {Cleland}},\ }\href {\doibase 10.1038/s41586-018-0719-5} {\bibfield
  {journal} {\bibinfo  {journal} {Nature}\ }\textbf {\bibinfo {volume} {563}},\
  \bibinfo {pages} {661} (\bibinfo {year} {2018})}\BibitemShut {NoStop}%
\bibitem [{\citenamefont {Chu}\ \emph {et~al.}(2018)\citenamefont {Chu},
  \citenamefont {Kharel}, \citenamefont {Yoon}, \citenamefont {Frunzio},
  \citenamefont {Rakich},\ and\ \citenamefont {Schoelkopf}}]{chu18}%
  \BibitemOpen
  \bibfield  {author} {\bibinfo {author} {\bibfnamefont {Y.}~\bibnamefont
  {Chu}}, \bibinfo {author} {\bibfnamefont {P.}~\bibnamefont {Kharel}},
  \bibinfo {author} {\bibfnamefont {T.}~\bibnamefont {Yoon}}, \bibinfo {author}
  {\bibfnamefont {L.}~\bibnamefont {Frunzio}}, \bibinfo {author} {\bibfnamefont
  {P.~T.}\ \bibnamefont {Rakich}}, \ and\ \bibinfo {author} {\bibfnamefont
  {R.~J.}\ \bibnamefont {Schoelkopf}},\ }\href {\doibase
  10.1038/s41586-018-0717-7} {\bibfield  {journal} {\bibinfo  {journal}
  {Nature}\ }\textbf {\bibinfo {volume} {563}},\ \bibinfo {pages} {666}
  (\bibinfo {year} {2018})}\BibitemShut {NoStop}%
\bibitem [{\citenamefont {Halperin}\ and\ \citenamefont
  {Varoquaux}(1990)}]{hal90}%
  \BibitemOpen
  \bibfield  {author} {\bibinfo {author} {\bibfnamefont {W.}~\bibnamefont
  {Halperin}}\ and\ \bibinfo {author} {\bibfnamefont {E.}~\bibnamefont
  {Varoquaux}},\ }in\ \href@noop {} {\emph {\bibinfo {booktitle} {Helium Three,
  Modern Problems in Condensed Matter Sciences}}},\ Vol.~\bibinfo {volume}
  {26},\ \bibinfo {editor} {edited by\ \bibinfo {editor} {\bibfnamefont
  {W.}~\bibnamefont {Halperin}}\ and\ \bibinfo {editor} {\bibfnamefont
  {L.}~\bibnamefont {Pitaevskii}}}\ (\bibinfo  {publisher} {Elsevier BV},\
  \bibinfo {year} {1990})\ Chap.~\bibinfo {chapter} {7}, pp.\ \bibinfo {pages}
  {353--522}\BibitemShut {NoStop}%
\bibitem [{\citenamefont {Davis}\ \emph {et~al.}(2008)\citenamefont {Davis},
  \citenamefont {Pollanen}, \citenamefont {Choi}, \citenamefont {Sauls},\ and\
  \citenamefont {Halperin}}]{Dav08}%
  \BibitemOpen
  \bibfield  {author} {\bibinfo {author} {\bibfnamefont {J.~P.}\ \bibnamefont
  {Davis}}, \bibinfo {author} {\bibfnamefont {J.}~\bibnamefont {Pollanen}},
  \bibinfo {author} {\bibfnamefont {H.}~\bibnamefont {Choi}}, \bibinfo {author}
  {\bibfnamefont {J.~A.}\ \bibnamefont {Sauls}}, \ and\ \bibinfo {author}
  {\bibfnamefont {W.~P.}\ \bibnamefont {Halperin}},\ }\href {\doibase
  10.1038/nphys969} {\bibfield  {journal} {\bibinfo  {journal} {Nature
  Physics}\ }\textbf {\bibinfo {volume} {4}},\ \bibinfo {pages} {571} (\bibinfo
  {year} {2008})}\BibitemShut {NoStop}%
\bibitem [{\citenamefont {Koch}\ \emph {et~al.}(2007)\citenamefont {Koch},
  \citenamefont {Yu}, \citenamefont {Gambetta}, \citenamefont {Houck},
  \citenamefont {Schuster}, \citenamefont {Majer}, \citenamefont {Blais},
  \citenamefont {Devoret}, \citenamefont {Girvin},\ and\ \citenamefont
  {Schoelkopf}}]{koc07}%
  \BibitemOpen
  \bibfield  {author} {\bibinfo {author} {\bibfnamefont {J.}~\bibnamefont
  {Koch}}, \bibinfo {author} {\bibfnamefont {T.~M.}\ \bibnamefont {Yu}},
  \bibinfo {author} {\bibfnamefont {J.}~\bibnamefont {Gambetta}}, \bibinfo
  {author} {\bibfnamefont {A.~A.}\ \bibnamefont {Houck}}, \bibinfo {author}
  {\bibfnamefont {D.~I.}\ \bibnamefont {Schuster}}, \bibinfo {author}
  {\bibfnamefont {J.}~\bibnamefont {Majer}}, \bibinfo {author} {\bibfnamefont
  {A.}~\bibnamefont {Blais}}, \bibinfo {author} {\bibfnamefont {M.~H.}\
  \bibnamefont {Devoret}}, \bibinfo {author} {\bibfnamefont {S.~M.}\
  \bibnamefont {Girvin}}, \ and\ \bibinfo {author} {\bibfnamefont {R.~J.}\
  \bibnamefont {Schoelkopf}},\ }\href {\doibase 10.1103/PhysRevA.76.042319}
  {\bibfield  {journal} {\bibinfo  {journal} {Phys. Rev. A}\ }\textbf {\bibinfo
  {volume} {76}},\ \bibinfo {pages} {042319} (\bibinfo {year}
  {2007})}\BibitemShut {NoStop}%
\bibitem [{Note3()}]{Note3}%
  \BibitemOpen
  \bibinfo {note} {Gilbert Engineering part \# 0119-783-1}\BibitemShut
  {NoStop}%
\bibitem [{\citenamefont {Fragner}(2013)}]{Fra13}%
  \BibitemOpen
  \bibfield  {author} {\bibinfo {author} {\bibfnamefont {A.}~\bibnamefont
  {Fragner}},\ }\emph {\bibinfo {title} {Circuit Quantum Electrodynamics with
  Electrons on Helium}},\ \href@noop {} {Ph.D. thesis},\ \bibinfo  {school}
  {Yale University} (\bibinfo {year} {2013})\BibitemShut {NoStop}%
\bibitem [{\citenamefont {Boissonneault}\ \emph {et~al.}(2010)\citenamefont
  {Boissonneault}, \citenamefont {Gambetta},\ and\ \citenamefont
  {Blais}}]{boi10}%
  \BibitemOpen
  \bibfield  {author} {\bibinfo {author} {\bibfnamefont {M.}~\bibnamefont
  {Boissonneault}}, \bibinfo {author} {\bibfnamefont {J.~M.}\ \bibnamefont
  {Gambetta}}, \ and\ \bibinfo {author} {\bibfnamefont {A.}~\bibnamefont
  {Blais}},\ }\href {\doibase 10.1103/PhysRevLett.105.100504} {\bibfield
  {journal} {\bibinfo  {journal} {Phys. Rev. Lett.}\ }\textbf {\bibinfo
  {volume} {105}},\ \bibinfo {pages} {100504} (\bibinfo {year}
  {2010})}\BibitemShut {NoStop}%
\bibitem [{\citenamefont {Bishop}\ \emph {et~al.}(2010)\citenamefont {Bishop},
  \citenamefont {Ginossar},\ and\ \citenamefont {Girvin}}]{bis10}%
  \BibitemOpen
  \bibfield  {author} {\bibinfo {author} {\bibfnamefont {L.~S.}\ \bibnamefont
  {Bishop}}, \bibinfo {author} {\bibfnamefont {E.}~\bibnamefont {Ginossar}}, \
  and\ \bibinfo {author} {\bibfnamefont {S.~M.}\ \bibnamefont {Girvin}},\
  }\href {\doibase 10.1103/PhysRevLett.105.100505} {\bibfield  {journal}
  {\bibinfo  {journal} {Phys. Rev. Lett.}\ }\textbf {\bibinfo {volume} {105}},\
  \bibinfo {pages} {100505} (\bibinfo {year} {2010})}\BibitemShut {NoStop}%
\bibitem [{\citenamefont {Reed}\ \emph {et~al.}(2010)\citenamefont {Reed},
  \citenamefont {DiCarlo}, \citenamefont {Johnson}, \citenamefont {Sun},
  \citenamefont {Schuster}, \citenamefont {Frunzio},\ and\ \citenamefont
  {Schoelkopf}}]{ree10}%
  \BibitemOpen
  \bibfield  {author} {\bibinfo {author} {\bibfnamefont {M.~D.}\ \bibnamefont
  {Reed}}, \bibinfo {author} {\bibfnamefont {L.}~\bibnamefont {DiCarlo}},
  \bibinfo {author} {\bibfnamefont {B.~R.}\ \bibnamefont {Johnson}}, \bibinfo
  {author} {\bibfnamefont {L.}~\bibnamefont {Sun}}, \bibinfo {author}
  {\bibfnamefont {D.~I.}\ \bibnamefont {Schuster}}, \bibinfo {author}
  {\bibfnamefont {L.}~\bibnamefont {Frunzio}}, \ and\ \bibinfo {author}
  {\bibfnamefont {R.~J.}\ \bibnamefont {Schoelkopf}},\ }\href {\doibase
  10.1103/PhysRevLett.105.173601} {\bibfield  {journal} {\bibinfo  {journal}
  {Phys. Rev. Lett.}\ }\textbf {\bibinfo {volume} {105}},\ \bibinfo {pages}
  {173601} (\bibinfo {year} {2010})}\BibitemShut {NoStop}%
\bibitem [{\citenamefont {Brooks}\ and\ \citenamefont
  {Donnelly}(1977)}]{bro77}%
  \BibitemOpen
  \bibfield  {author} {\bibinfo {author} {\bibfnamefont {J.~S.}\ \bibnamefont
  {Brooks}}\ and\ \bibinfo {author} {\bibfnamefont {R.~J.}\ \bibnamefont
  {Donnelly}},\ }\href {\doibase 10.1063/1.555549} {\bibfield  {journal}
  {\bibinfo  {journal} {Journal of Physical and Chemical Reference Data}\
  }\textbf {\bibinfo {volume} {6}},\ \bibinfo {pages} {51} (\bibinfo {year}
  {1977})},\ \Eprint {http://arxiv.org/abs/https://doi.org/10.1063/1.555549}
  {https://doi.org/10.1063/1.555549} \BibitemShut {NoStop}%
\bibitem [{\citenamefont {Schuster}\ \emph {et~al.}(2005)\citenamefont
  {Schuster}, \citenamefont {Wallraff}, \citenamefont {Blais}, \citenamefont
  {Frunzio}, \citenamefont {Huang}, \citenamefont {Majer}, \citenamefont
  {Girvin},\ and\ \citenamefont {Schoelkopf}}]{schu05}%
  \BibitemOpen
  \bibfield  {author} {\bibinfo {author} {\bibfnamefont {D.~I.}\ \bibnamefont
  {Schuster}}, \bibinfo {author} {\bibfnamefont {A.}~\bibnamefont {Wallraff}},
  \bibinfo {author} {\bibfnamefont {A.}~\bibnamefont {Blais}}, \bibinfo
  {author} {\bibfnamefont {L.}~\bibnamefont {Frunzio}}, \bibinfo {author}
  {\bibfnamefont {R.-S.}\ \bibnamefont {Huang}}, \bibinfo {author}
  {\bibfnamefont {J.}~\bibnamefont {Majer}}, \bibinfo {author} {\bibfnamefont
  {S.~M.}\ \bibnamefont {Girvin}}, \ and\ \bibinfo {author} {\bibfnamefont
  {R.~J.}\ \bibnamefont {Schoelkopf}},\ }\href {\doibase
  10.1103/PhysRevLett.94.123602} {\bibfield  {journal} {\bibinfo  {journal}
  {Phys. Rev. Lett.}\ }\textbf {\bibinfo {volume} {94}},\ \bibinfo {pages}
  {123602} (\bibinfo {year} {2005})}\BibitemShut {NoStop}%
\bibitem [{\citenamefont {Rogers}\ and\ \citenamefont {Buhrman}(1985)}]{rog85}%
  \BibitemOpen
  \bibfield  {author} {\bibinfo {author} {\bibfnamefont {C.~T.}\ \bibnamefont
  {Rogers}}\ and\ \bibinfo {author} {\bibfnamefont {R.~A.}\ \bibnamefont
  {Buhrman}},\ }\href {\doibase 10.1103/PhysRevLett.55.859} {\bibfield
  {journal} {\bibinfo  {journal} {Phys. Rev. Lett.}\ }\textbf {\bibinfo
  {volume} {55}},\ \bibinfo {pages} {859} (\bibinfo {year} {1985})}\BibitemShut
  {NoStop}%
\bibitem [{\citenamefont {Van~Harlingen}\ \emph {et~al.}(2004)\citenamefont
  {Van~Harlingen}, \citenamefont {Robertson}, \citenamefont {Plourde},
  \citenamefont {Reichardt}, \citenamefont {Crane},\ and\ \citenamefont
  {Clarke}}]{van04}%
  \BibitemOpen
  \bibfield  {author} {\bibinfo {author} {\bibfnamefont {D.~J.}\ \bibnamefont
  {Van~Harlingen}}, \bibinfo {author} {\bibfnamefont {T.~L.}\ \bibnamefont
  {Robertson}}, \bibinfo {author} {\bibfnamefont {B.~L.~T.}\ \bibnamefont
  {Plourde}}, \bibinfo {author} {\bibfnamefont {P.~A.}\ \bibnamefont
  {Reichardt}}, \bibinfo {author} {\bibfnamefont {T.~A.}\ \bibnamefont
  {Crane}}, \ and\ \bibinfo {author} {\bibfnamefont {J.}~\bibnamefont
  {Clarke}},\ }\href {\doibase 10.1103/PhysRevB.70.064517} {\bibfield
  {journal} {\bibinfo  {journal} {Phys. Rev. B}\ }\textbf {\bibinfo {volume}
  {70}},\ \bibinfo {pages} {064517} (\bibinfo {year} {2004})}\BibitemShut
  {NoStop}%
\bibitem [{Note4()}]{Note4}%
  \BibitemOpen
  \bibinfo {note} {We note that at the large qubit/cavity detuning used in this
  experiment, the increase in the Purcell emission rate $\Gamma _p$ caused by
  the helium induced shift of the cavity frequency is negligible compared to
  the long timescale fluctuations in the emission rate. Specifically $\Gamma _p
  = (g_{01}/\Delta )^2 \kappa \sim (265~\mu $s$)^{-1}$ for the empty cavity and
  $\sim (240~\mu $s$)^{-1}$ for the full cavity, where $\kappa $ is the cavity
  linewidth \cite {koc07}.}\BibitemShut {Stop}%
\bibitem [{\citenamefont {Jin}\ \emph {et~al.}(2015)\citenamefont {Jin},
  \citenamefont {Kamal}, \citenamefont {Sears}, \citenamefont {Gudmundsen},
  \citenamefont {Hover}, \citenamefont {Miloshi}, \citenamefont {Slattery},
  \citenamefont {Yan}, \citenamefont {Yoder}, \citenamefont {Orlando},
  \citenamefont {Gustavsson},\ and\ \citenamefont {Oliver}}]{jin15}%
  \BibitemOpen
  \bibfield  {author} {\bibinfo {author} {\bibfnamefont {X.~Y.}\ \bibnamefont
  {Jin}}, \bibinfo {author} {\bibfnamefont {A.}~\bibnamefont {Kamal}}, \bibinfo
  {author} {\bibfnamefont {A.~P.}\ \bibnamefont {Sears}}, \bibinfo {author}
  {\bibfnamefont {T.}~\bibnamefont {Gudmundsen}}, \bibinfo {author}
  {\bibfnamefont {D.}~\bibnamefont {Hover}}, \bibinfo {author} {\bibfnamefont
  {J.}~\bibnamefont {Miloshi}}, \bibinfo {author} {\bibfnamefont
  {R.}~\bibnamefont {Slattery}}, \bibinfo {author} {\bibfnamefont
  {F.}~\bibnamefont {Yan}}, \bibinfo {author} {\bibfnamefont {J.}~\bibnamefont
  {Yoder}}, \bibinfo {author} {\bibfnamefont {T.~P.}\ \bibnamefont {Orlando}},
  \bibinfo {author} {\bibfnamefont {S.}~\bibnamefont {Gustavsson}}, \ and\
  \bibinfo {author} {\bibfnamefont {W.~D.}\ \bibnamefont {Oliver}},\ }\href
  {\doibase 10.1103/PhysRevLett.114.240501} {\bibfield  {journal} {\bibinfo
  {journal} {Phys. Rev. Lett.}\ }\textbf {\bibinfo {volume} {114}},\ \bibinfo
  {pages} {240501} (\bibinfo {year} {2015})}\BibitemShut {NoStop}%
\bibitem [{\citenamefont {Pobell}(2007)}]{Pob92}%
  \BibitemOpen
  \bibfield  {author} {\bibinfo {author} {\bibfnamefont {F.}~\bibnamefont
  {Pobell}},\ }\href@noop {} {\emph {\bibinfo {title} {Matter and Methods at
  Low Temperatures}}},\ \bibinfo {edition} {3rd}\ ed.\ (\bibinfo  {publisher}
  {Springer, Berlin},\ \bibinfo {year} {2007})\BibitemShut {NoStop}%
\bibitem [{\citenamefont {Pollanen}\ \emph {et~al.}(2009)\citenamefont
  {Pollanen}, \citenamefont {Choi}, \citenamefont {Davis}, \citenamefont
  {Rolfs},\ and\ \citenamefont {Halperin}}]{Pol09}%
  \BibitemOpen
  \bibfield  {author} {\bibinfo {author} {\bibfnamefont {J.}~\bibnamefont
  {Pollanen}}, \bibinfo {author} {\bibfnamefont {H.}~\bibnamefont {Choi}},
  \bibinfo {author} {\bibfnamefont {J.}~\bibnamefont {Davis}}, \bibinfo
  {author} {\bibfnamefont {B.}~\bibnamefont {Rolfs}}, \ and\ \bibinfo {author}
  {\bibfnamefont {W.}~\bibnamefont {Halperin}},\ }\href@noop {} {\bibfield
  {journal} {\bibinfo  {journal} {J. Phys. Con. Ser.}\ }\textbf {\bibinfo
  {volume} {150}},\ \bibinfo {pages} {012037} (\bibinfo {year}
  {2009})}\BibitemShut {NoStop}%
\bibitem [{\citenamefont {M\"uller}\ \emph {et~al.}(2015)\citenamefont
  {M\"uller}, \citenamefont {Lisenfeld}, \citenamefont {Shnirman},\ and\
  \citenamefont {Poletto}}]{mul15}%
  \BibitemOpen
  \bibfield  {author} {\bibinfo {author} {\bibfnamefont {C.}~\bibnamefont
  {M\"uller}}, \bibinfo {author} {\bibfnamefont {J.}~\bibnamefont {Lisenfeld}},
  \bibinfo {author} {\bibfnamefont {A.}~\bibnamefont {Shnirman}}, \ and\
  \bibinfo {author} {\bibfnamefont {S.}~\bibnamefont {Poletto}},\ }\href
  {\doibase 10.1103/PhysRevB.92.035442} {\bibfield  {journal} {\bibinfo
  {journal} {Phys. Rev. B}\ }\textbf {\bibinfo {volume} {92}},\ \bibinfo
  {pages} {035442} (\bibinfo {year} {2015})}\BibitemShut {NoStop}%
\bibitem [{\citenamefont {Burnett}\ \emph {et~al.}(2019)\citenamefont
  {Burnett}, \citenamefont {Bengtsson}, \citenamefont {Scigliuzzo},
  \citenamefont {Niepce}, \citenamefont {Kudra}, \citenamefont {Delsing},\ and\
  \citenamefont {Bylander}}]{bur19}%
  \BibitemOpen
  \bibfield  {author} {\bibinfo {author} {\bibfnamefont {J.~J.}\ \bibnamefont
  {Burnett}}, \bibinfo {author} {\bibfnamefont {A.}~\bibnamefont {Bengtsson}},
  \bibinfo {author} {\bibfnamefont {M.}~\bibnamefont {Scigliuzzo}}, \bibinfo
  {author} {\bibfnamefont {D.}~\bibnamefont {Niepce}}, \bibinfo {author}
  {\bibfnamefont {M.}~\bibnamefont {Kudra}}, \bibinfo {author} {\bibfnamefont
  {P.}~\bibnamefont {Delsing}}, \ and\ \bibinfo {author} {\bibfnamefont
  {J.}~\bibnamefont {Bylander}},\ }\href {\doibase 10.1038/s41534-019-0168-5}
  {\bibfield  {journal} {\bibinfo  {journal} {npj Quantum Information}\
  }\textbf {\bibinfo {volume} {5}},\ \bibinfo {pages} {54} (\bibinfo {year}
  {2019})}\BibitemShut {NoStop}%
\bibitem [{\citenamefont {Schl{\"o}r}\ \emph {et~al.}(2019)\citenamefont
  {Schl{\"o}r}, \citenamefont {Lisenfeld}, \citenamefont {M{\"u}ller},
  \citenamefont {Bilmes}, \citenamefont {Schneider}, \citenamefont {Pappas},
  \citenamefont {Ustinov},\ and\ \citenamefont {Weides}}]{sch19}%
  \BibitemOpen
  \bibfield  {author} {\bibinfo {author} {\bibfnamefont {S.}~\bibnamefont
  {Schl{\"o}r}}, \bibinfo {author} {\bibfnamefont {J.}~\bibnamefont
  {Lisenfeld}}, \bibinfo {author} {\bibfnamefont {C.}~\bibnamefont
  {M{\"u}ller}}, \bibinfo {author} {\bibfnamefont {A.}~\bibnamefont {Bilmes}},
  \bibinfo {author} {\bibfnamefont {A.}~\bibnamefont {Schneider}}, \bibinfo
  {author} {\bibfnamefont {D.~P.}\ \bibnamefont {Pappas}}, \bibinfo {author}
  {\bibfnamefont {A.~V.}\ \bibnamefont {Ustinov}}, \ and\ \bibinfo {author}
  {\bibfnamefont {M.}~\bibnamefont {Weides}},\ }\href@noop {} {\bibfield
  {journal} {\bibinfo  {journal} {arXiv:1901.05352v2}\ } (\bibinfo {year}
  {2019})}\BibitemShut {NoStop}%
\bibitem [{\citenamefont {Yan}\ \emph {et~al.}(2016)\citenamefont {Yan},
  \citenamefont {Gustavsson}, \citenamefont {Kamal}, \citenamefont {Birenbaum},
  \citenamefont {Sears}, \citenamefont {Hover}, \citenamefont {Gudmundsen},
  \citenamefont {Rosenberg}, \citenamefont {Samach}, \citenamefont {Weber},
  \citenamefont {Yoder}, \citenamefont {Orlando}, \citenamefont {Clarke},
  \citenamefont {Kerman},\ and\ \citenamefont {Oliver}}]{yan16}%
  \BibitemOpen
  \bibfield  {author} {\bibinfo {author} {\bibfnamefont {F.}~\bibnamefont
  {Yan}}, \bibinfo {author} {\bibfnamefont {S.}~\bibnamefont {Gustavsson}},
  \bibinfo {author} {\bibfnamefont {A.}~\bibnamefont {Kamal}}, \bibinfo
  {author} {\bibfnamefont {J.}~\bibnamefont {Birenbaum}}, \bibinfo {author}
  {\bibfnamefont {A.~P.}\ \bibnamefont {Sears}}, \bibinfo {author}
  {\bibfnamefont {D.}~\bibnamefont {Hover}}, \bibinfo {author} {\bibfnamefont
  {T.~J.}\ \bibnamefont {Gudmundsen}}, \bibinfo {author} {\bibfnamefont
  {D.}~\bibnamefont {Rosenberg}}, \bibinfo {author} {\bibfnamefont
  {G.}~\bibnamefont {Samach}}, \bibinfo {author} {\bibfnamefont
  {S.}~\bibnamefont {Weber}}, \bibinfo {author} {\bibfnamefont {J.~L.}\
  \bibnamefont {Yoder}}, \bibinfo {author} {\bibfnamefont {T.~P.}\ \bibnamefont
  {Orlando}}, \bibinfo {author} {\bibfnamefont {J.}~\bibnamefont {Clarke}},
  \bibinfo {author} {\bibfnamefont {A.~J.}\ \bibnamefont {Kerman}}, \ and\
  \bibinfo {author} {\bibfnamefont {W.~D.}\ \bibnamefont {Oliver}},\ }\href
  {http://dx.doi.org/10.1038/ncomms12964} {\bibfield  {journal} {\bibinfo
  {journal} {Nature Communications}\ }\textbf {\bibinfo {volume} {7}},\
  \bibinfo {pages} {12964 EP } (\bibinfo {year} {2016})}\BibitemShut {NoStop}%
\bibitem [{\citenamefont {Sears}\ \emph {et~al.}(2012)\citenamefont {Sears},
  \citenamefont {Petrenko}, \citenamefont {Catelani}, \citenamefont {Sun},
  \citenamefont {Paik}, \citenamefont {Kirchmair}, \citenamefont {Frunzio},
  \citenamefont {Glazman}, \citenamefont {Girvin},\ and\ \citenamefont
  {Schoelkopf}}]{sea12}%
  \BibitemOpen
  \bibfield  {author} {\bibinfo {author} {\bibfnamefont {A.~P.}\ \bibnamefont
  {Sears}}, \bibinfo {author} {\bibfnamefont {A.}~\bibnamefont {Petrenko}},
  \bibinfo {author} {\bibfnamefont {G.}~\bibnamefont {Catelani}}, \bibinfo
  {author} {\bibfnamefont {L.}~\bibnamefont {Sun}}, \bibinfo {author}
  {\bibfnamefont {H.}~\bibnamefont {Paik}}, \bibinfo {author} {\bibfnamefont
  {G.}~\bibnamefont {Kirchmair}}, \bibinfo {author} {\bibfnamefont
  {L.}~\bibnamefont {Frunzio}}, \bibinfo {author} {\bibfnamefont {L.~I.}\
  \bibnamefont {Glazman}}, \bibinfo {author} {\bibfnamefont {S.~M.}\
  \bibnamefont {Girvin}}, \ and\ \bibinfo {author} {\bibfnamefont {R.~J.}\
  \bibnamefont {Schoelkopf}},\ }\href {\doibase 10.1103/PhysRevB.86.180504}
  {\bibfield  {journal} {\bibinfo  {journal} {Phys. Rev. B}\ }\textbf {\bibinfo
  {volume} {86}},\ \bibinfo {pages} {180504} (\bibinfo {year}
  {2012})}\BibitemShut {NoStop}%
\bibitem [{\citenamefont {Clerk}\ and\ \citenamefont {Utami}(2007)}]{cle07}%
  \BibitemOpen
  \bibfield  {author} {\bibinfo {author} {\bibfnamefont {A.~A.}\ \bibnamefont
  {Clerk}}\ and\ \bibinfo {author} {\bibfnamefont {D.~W.}\ \bibnamefont
  {Utami}},\ }\href {\doibase 10.1103/PhysRevA.75.042302} {\bibfield  {journal}
  {\bibinfo  {journal} {Phys. Rev. A}\ }\textbf {\bibinfo {volume} {75}},\
  \bibinfo {pages} {042302} (\bibinfo {year} {2007})}\BibitemShut {NoStop}%
\bibitem [{\citenamefont {Geerlings}\ \emph {et~al.}(2013)\citenamefont
  {Geerlings}, \citenamefont {Leghtas}, \citenamefont {Pop}, \citenamefont
  {Shankar}, \citenamefont {Frunzio}, \citenamefont {Schoelkopf}, \citenamefont
  {Mirrahimi},\ and\ \citenamefont {Devoret}}]{gee13}%
  \BibitemOpen
  \bibfield  {author} {\bibinfo {author} {\bibfnamefont {K.}~\bibnamefont
  {Geerlings}}, \bibinfo {author} {\bibfnamefont {Z.}~\bibnamefont {Leghtas}},
  \bibinfo {author} {\bibfnamefont {I.~M.}\ \bibnamefont {Pop}}, \bibinfo
  {author} {\bibfnamefont {S.}~\bibnamefont {Shankar}}, \bibinfo {author}
  {\bibfnamefont {L.}~\bibnamefont {Frunzio}}, \bibinfo {author} {\bibfnamefont
  {R.~J.}\ \bibnamefont {Schoelkopf}}, \bibinfo {author} {\bibfnamefont
  {M.}~\bibnamefont {Mirrahimi}}, \ and\ \bibinfo {author} {\bibfnamefont
  {M.~H.}\ \bibnamefont {Devoret}},\ }\href {\doibase
  10.1103/PhysRevLett.110.120501} {\bibfield  {journal} {\bibinfo  {journal}
  {Phys. Rev. Lett.}\ }\textbf {\bibinfo {volume} {110}},\ \bibinfo {pages}
  {120501} (\bibinfo {year} {2013})}\BibitemShut {NoStop}%
\bibitem [{\citenamefont {Klimov}\ \emph {et~al.}(2018)\citenamefont {Klimov},
  \citenamefont {Kelly}, \citenamefont {Chen}, \citenamefont {Neeley},
  \citenamefont {Megrant}, \citenamefont {Burkett}, \citenamefont {Barends},
  \citenamefont {Arya}, \citenamefont {Chiaro}, \citenamefont {Chen},
  \citenamefont {Dunsworth}, \citenamefont {Fowler}, \citenamefont {Foxen},
  \citenamefont {Gidney}, \citenamefont {Giustina}, \citenamefont {Graff},
  \citenamefont {Huang}, \citenamefont {Jeffrey}, \citenamefont {Lucero},
  \citenamefont {Mutus}, \citenamefont {Naaman}, \citenamefont {Neill},
  \citenamefont {Quintana}, \citenamefont {Roushan}, \citenamefont {Sank},
  \citenamefont {Vainsencher}, \citenamefont {Wenner}, \citenamefont {White},
  \citenamefont {Boixo}, \citenamefont {Babbush}, \citenamefont {Smelyanskiy},
  \citenamefont {Neven}, ,\ and\ \citenamefont {Martinis}}]{kli18}%
  \BibitemOpen
  \bibfield  {author} {\bibinfo {author} {\bibfnamefont {P.}~\bibnamefont
  {Klimov}}, \bibinfo {author} {\bibfnamefont {J.}~\bibnamefont {Kelly}},
  \bibinfo {author} {\bibfnamefont {Z.}~\bibnamefont {Chen}}, \bibinfo {author}
  {\bibfnamefont {M.}~\bibnamefont {Neeley}}, \bibinfo {author} {\bibfnamefont
  {A.}~\bibnamefont {Megrant}}, \bibinfo {author} {\bibfnamefont
  {B.}~\bibnamefont {Burkett}}, \bibinfo {author} {\bibfnamefont
  {R.}~\bibnamefont {Barends}}, \bibinfo {author} {\bibfnamefont
  {K.}~\bibnamefont {Arya}}, \bibinfo {author} {\bibfnamefont {B.}~\bibnamefont
  {Chiaro}}, \bibinfo {author} {\bibfnamefont {Y.}~\bibnamefont {Chen}},
  \bibinfo {author} {\bibfnamefont {A.}~\bibnamefont {Dunsworth}}, \bibinfo
  {author} {\bibfnamefont {A.}~\bibnamefont {Fowler}}, \bibinfo {author}
  {\bibfnamefont {B.}~\bibnamefont {Foxen}}, \bibinfo {author} {\bibfnamefont
  {C.}~\bibnamefont {Gidney}}, \bibinfo {author} {\bibfnamefont
  {M.}~\bibnamefont {Giustina}}, \bibinfo {author} {\bibfnamefont
  {R.}~\bibnamefont {Graff}}, \bibinfo {author} {\bibfnamefont
  {T.}~\bibnamefont {Huang}}, \bibinfo {author} {\bibfnamefont
  {E.}~\bibnamefont {Jeffrey}}, \bibinfo {author} {\bibfnamefont
  {E.}~\bibnamefont {Lucero}}, \bibinfo {author} {\bibfnamefont
  {J.}~\bibnamefont {Mutus}}, \bibinfo {author} {\bibfnamefont
  {O.}~\bibnamefont {Naaman}}, \bibinfo {author} {\bibfnamefont
  {C.}~\bibnamefont {Neill}}, \bibinfo {author} {\bibfnamefont
  {C.}~\bibnamefont {Quintana}}, \bibinfo {author} {\bibfnamefont
  {P.}~\bibnamefont {Roushan}}, \bibinfo {author} {\bibfnamefont
  {D.}~\bibnamefont {Sank}}, \bibinfo {author} {\bibfnamefont {A.}~\bibnamefont
  {Vainsencher}}, \bibinfo {author} {\bibfnamefont {J.}~\bibnamefont {Wenner}},
  \bibinfo {author} {\bibfnamefont {T.~C.}\ \bibnamefont {White}}, \bibinfo
  {author} {\bibfnamefont {S.}~\bibnamefont {Boixo}}, \bibinfo {author}
  {\bibfnamefont {R.}~\bibnamefont {Babbush}}, \bibinfo {author} {\bibfnamefont
  {V.}~\bibnamefont {Smelyanskiy}}, \bibinfo {author} {\bibfnamefont
  {H.}~\bibnamefont {Neven}}, , \ and\ \bibinfo {author} {\bibfnamefont
  {J.~M.}\ \bibnamefont {Martinis}},\ }\href@noop {} {\bibfield  {journal}
  {\bibinfo  {journal} {Phys. Rev. Lett.}\ }\textbf {\bibinfo {volume} {121}},\
  \bibinfo {pages} {090502} (\bibinfo {year} {2018})}\BibitemShut {NoStop}%
\bibitem [{\citenamefont {Everitt}\ \emph {et~al.}(1964)\citenamefont
  {Everitt}, \citenamefont {Atkins},\ and\ \citenamefont {Denenstein}}]{Eve64}%
  \BibitemOpen
  \bibfield  {author} {\bibinfo {author} {\bibfnamefont {C.~W.~F.}\
  \bibnamefont {Everitt}}, \bibinfo {author} {\bibfnamefont {K.~R.}\
  \bibnamefont {Atkins}}, \ and\ \bibinfo {author} {\bibfnamefont
  {A.}~\bibnamefont {Denenstein}},\ }\href {\doibase 10.1103/PhysRev.136.A1494}
  {\bibfield  {journal} {\bibinfo  {journal} {Phys. Rev.}\ }\textbf {\bibinfo
  {volume} {136}},\ \bibinfo {pages} {A1494} (\bibinfo {year}
  {1964})}\BibitemShut {NoStop}%
\bibitem [{\citenamefont {Schechter}\ \emph {et~al.}(1998)\citenamefont
  {Schechter}, \citenamefont {Simmonds}, \citenamefont {Packard},\ and\
  \citenamefont {Davis}}]{Sch98}%
  \BibitemOpen
  \bibfield  {author} {\bibinfo {author} {\bibfnamefont {A.~M.~R.}\
  \bibnamefont {Schechter}}, \bibinfo {author} {\bibfnamefont {R.~W.}\
  \bibnamefont {Simmonds}}, \bibinfo {author} {\bibfnamefont {R.~E.}\
  \bibnamefont {Packard}}, \ and\ \bibinfo {author} {\bibfnamefont {J.~C.}\
  \bibnamefont {Davis}},\ }\href {\doibase 10.1038/25090} {\bibfield  {journal}
  {\bibinfo  {journal} {Nature}\ }\textbf {\bibinfo {volume} {396}},\ \bibinfo
  {pages} {554} (\bibinfo {year} {1998})}\BibitemShut {NoStop}%
\bibitem [{\citenamefont {Saam}(1975)}]{Saa75}%
  \BibitemOpen
  \bibfield  {author} {\bibinfo {author} {\bibfnamefont {W.~F.}\ \bibnamefont
  {Saam}},\ }\href {\doibase 10.1103/PhysRevB.12.163} {\bibfield  {journal}
  {\bibinfo  {journal} {Phys. Rev. B}\ }\textbf {\bibinfo {volume} {12}},\
  \bibinfo {pages} {163} (\bibinfo {year} {1975})}\BibitemShut {NoStop}%
\bibitem [{\citenamefont {Dykman}\ \emph {et~al.}(2017)\citenamefont {Dykman},
  \citenamefont {Kono}, \citenamefont {Konstantinov},\ and\ \citenamefont
  {Lea}}]{dyk17}%
  \BibitemOpen
  \bibfield  {author} {\bibinfo {author} {\bibfnamefont {M.}~\bibnamefont
  {Dykman}}, \bibinfo {author} {\bibfnamefont {K.}~\bibnamefont {Kono}},
  \bibinfo {author} {\bibfnamefont {D.}~\bibnamefont {Konstantinov}}, \ and\
  \bibinfo {author} {\bibfnamefont {M.}~\bibnamefont {Lea}},\ }\href@noop {}
  {\bibfield  {journal} {\bibinfo  {journal} {Phys. Rev. Lett.}\ }\textbf
  {\bibinfo {volume} {119}},\ \bibinfo {pages} {256802} (\bibinfo {year}
  {2017})}\BibitemShut {NoStop}%
\bibitem [{\citenamefont {Devoret}(1997)}]{dev97}%
  \BibitemOpen
  \bibfield  {author} {\bibinfo {author} {\bibfnamefont {M.~H.}\ \bibnamefont
  {Devoret}},\ }\enquote {\bibinfo {title} {Quantum fluctuations in electrical
  circuits},}\ \ (\bibinfo  {publisher} {Elsevier Science},\ \bibinfo {year}
  {1997})\ Chap.~\bibinfo {chapter} {10}\BibitemShut {NoStop}%
\bibitem [{Note5()}]{Note5}%
  \BibitemOpen
  \bibinfo {note} {This capacitance includes both the intrinsic Josephson
  junction capacitance and the shunt capacitance provided by the antenna
  paddles of the qubit.}\BibitemShut {Stop}%
\bibitem [{\citenamefont {Catelani}\ \emph {et~al.}(2012)\citenamefont
  {Catelani}, \citenamefont {Nigg}, \citenamefont {Girvin}, \citenamefont
  {Schoelkopf},\ and\ \citenamefont {Glazman}}]{cat12}%
  \BibitemOpen
  \bibfield  {author} {\bibinfo {author} {\bibfnamefont {G.}~\bibnamefont
  {Catelani}}, \bibinfo {author} {\bibfnamefont {S.~E.}\ \bibnamefont {Nigg}},
  \bibinfo {author} {\bibfnamefont {S.~M.}\ \bibnamefont {Girvin}}, \bibinfo
  {author} {\bibfnamefont {R.~J.}\ \bibnamefont {Schoelkopf}}, \ and\ \bibinfo
  {author} {\bibfnamefont {L.~I.}\ \bibnamefont {Glazman}},\ }\href {\doibase
  10.1103/PhysRevB.86.184514} {\bibfield  {journal} {\bibinfo  {journal} {Phys.
  Rev. B}\ }\textbf {\bibinfo {volume} {86}},\ \bibinfo {pages} {184514}
  (\bibinfo {year} {2012})}\BibitemShut {NoStop}%
\bibitem [{\citenamefont {Palacios-Laloy}\ \emph {et~al.}(2008)\citenamefont
  {Palacios-Laloy}, \citenamefont {Nguyen}, \citenamefont {abnd P.~Bertet},
  \citenamefont {Vion},\ and\ \citenamefont {Esteve}}]{Pal08}%
  \BibitemOpen
  \bibfield  {author} {\bibinfo {author} {\bibfnamefont {A.}~\bibnamefont
  {Palacios-Laloy}}, \bibinfo {author} {\bibfnamefont {F.}~\bibnamefont
  {Nguyen}}, \bibinfo {author} {\bibfnamefont {F.~M.}\ \bibnamefont {abnd
  P.~Bertet}}, \bibinfo {author} {\bibfnamefont {D.}~\bibnamefont {Vion}}, \
  and\ \bibinfo {author} {\bibfnamefont {D.}~\bibnamefont {Esteve}},\
  }\href@noop {} {\bibfield  {journal} {\bibinfo  {journal} {Journal of Low
  Temperature Physics}\ }\textbf {\bibinfo {volume} {151}},\ \bibinfo {pages}
  {1034} (\bibinfo {year} {2008})}\BibitemShut {NoStop}%
\bibitem [{\citenamefont {Sandberg}\ \emph {et~al.}(2008)\citenamefont
  {Sandberg}, \citenamefont {Wilson}, \citenamefont {Persson}, \citenamefont
  {Bauch}, \citenamefont {Johansson}, \citenamefont {Shumeiko}, \citenamefont
  {Duty},\ and\ \citenamefont {Delsing}}]{San08}%
  \BibitemOpen
  \bibfield  {author} {\bibinfo {author} {\bibfnamefont {M.}~\bibnamefont
  {Sandberg}}, \bibinfo {author} {\bibfnamefont {C.}~\bibnamefont {Wilson}},
  \bibinfo {author} {\bibfnamefont {F.}~\bibnamefont {Persson}}, \bibinfo
  {author} {\bibfnamefont {T.}~\bibnamefont {Bauch}}, \bibinfo {author}
  {\bibfnamefont {G.}~\bibnamefont {Johansson}}, \bibinfo {author}
  {\bibfnamefont {V.}~\bibnamefont {Shumeiko}}, \bibinfo {author}
  {\bibfnamefont {T.}~\bibnamefont {Duty}}, \ and\ \bibinfo {author}
  {\bibfnamefont {P.}~\bibnamefont {Delsing}},\ }\href@noop {} {\bibfield
  {journal} {\bibinfo  {journal} {Applied Physics Letters}\ }\textbf {\bibinfo
  {volume} {92}},\ \bibinfo {pages} {203501} (\bibinfo {year}
  {2008})}\BibitemShut {NoStop}%
\bibitem [{\citenamefont {Clark}\ \emph {et~al.}(2018)\citenamefont {Clark},
  \citenamefont {Vadakkumbatt}, \citenamefont {Souris}, \citenamefont {Ramp},\
  and\ \citenamefont {Davis}}]{cla18}%
  \BibitemOpen
  \bibfield  {author} {\bibinfo {author} {\bibfnamefont {T.}~\bibnamefont
  {Clark}}, \bibinfo {author} {\bibfnamefont {V.}~\bibnamefont {Vadakkumbatt}},
  \bibinfo {author} {\bibfnamefont {F.}~\bibnamefont {Souris}}, \bibinfo
  {author} {\bibfnamefont {H.}~\bibnamefont {Ramp}}, \ and\ \bibinfo {author}
  {\bibfnamefont {J.}~\bibnamefont {Davis}},\ }\href@noop {} {\bibfield
  {journal} {\bibinfo  {journal} {Rev. Sci. Instrum.}\ }\textbf {\bibinfo
  {volume} {89}},\ \bibinfo {pages} {114704} (\bibinfo {year}
  {2018})}\BibitemShut {NoStop}%
\end{thebibliography}

\end{document}